\documentclass[12pt,preprint]{aastex}
\newcommand\gsim{\,\lower3pt\hbox{$\sim$}\llap{\raise2pt\hbox{$>$}}\,}
\newcommand\lsim{\,\lower3pt\hbox{$\sim$}\llap{\raise2pt\hbox{$<$}}\,}

\newcommand{\bm}{\boldmath}
\newcommand{\ubm}{\unboldmath}
\newcommand{\be}{\begin{equation}}
\newcommand{\ee}{\end{equation}}

\newcommand{\tw}{{\mathcal T}}
\received{2005 January}
\begin{document}
\bibliographystyle{newapj}

\title{Magnetic Flux Tube Reconnection: Tunneling Versus Slingshot}

\author{M. G. Linton\altaffilmark{1} and S. K. Antiochos\altaffilmark{1}}
\authoraddr{Space Science Division, Naval Research Laboratory,
Washington, DC 20375-5352}
\altaffiltext{1} {Space Science Division, Naval Research Laboratory,
Washington, DC 20375-5352}

\begin{abstract}

The highly discrete nature of the solar magnetic field as it emerges 
into the corona through the photosphere, where it is predominantly 
concentrated into 
sunspots and magnetic pores, indicates that the magnetic field exists as 
discrete, isolated flux tubes in the convection zone, and will remain as 
discrete flux tubes in the corona until it collides and 
reconnects with other coronal fields. 
Collisions of these flux tubes, both in the convection zone and in the 
corona, will in general be three dimensional as the 
flux tubes will collide at random angles, and in many cases this will lead to
reconnection, both rearranging the magnetic field topology in fundamental
ways, and releasing magnetic energy.  With the goal of better understanding 
these dynamics, we carry out a set of numerical experiments 
exploring fundamental characteristics of three dimensional magnetic flux 
tube reconnection.  We first show that reconnecting flux tubes at 
opposite extremes of 
twist behave very differently: in some configurations, low twist tubes 
slingshot while high twist tubes tunnel. We then discuss a theory 
explaining these differences: by assuming helicity conservation during 
the reconnection one can show that at high twist, tunneled tubes reach a 
lower magnetic energy state than slingshot tubes, whereas at low twist 
the opposite holds.  We test three predictions made by this theory. 
1) We find that the level of twist at which the transition from slingshot 
to tunnel occurs is about two to three times higher than predicted on 
the basis of energetics and helicity conservation alone, probably because 
the dynamics of the reconnection play a large role as well.  2) We find 
that the tunnel occurs at all flux tube collision angles predicted by the 
theory. 3) We find that the amount of magnetic energy a slingshot or a 
tunnel reconnection releases agrees reasonably well with the theory, 
though at the high resistivities we have to use for numerical stability, 
a significant amount of magnetic energy is lost to diffusion, independent 
of reconnection.  We find that, while the slingshot reconnection is 
generally applicable to flux tubes of all twist, the level of twist 
needed for tunneling reconnection is relatively high compared to 
observations of the twist of large coronal loops. 
Therefore the tunnel is mainly relevant for the small scale, highly
twisted fields observed within large, slightly twisted sunspots
(Canfield, private communication) and in those convection zone flux
tubes which are so highly twisted that they can only emerge
partway into the corona \citep{Fan2001, Magara2001}.

\end{abstract}

\keywords{MHD---Sun: flares---Sun: magnetic fields}

\section{Introduction}

A fundamental question in the study of reconnection
is predicting the dynamics of two interacting, three
dimensional magnetic elements. In the context of solar physics, this
is important for understanding colliding coronal loops, the emergence
of sunspot active regions into pre-existing coronal field, and the
collision and reconnection of the flux elements of the magnetic carpet 
\citep[see e.g.][]{Schrijver1998, Priest2002}.
These reconnections are thought to play a major role in the dynamics of
solar flares, coronal mass ejections, and coronal heating 
\citep[see e.g.][]{Gold1960,Shibata1995,
Gosling1975,Mikic1994,Antiochos1999,Parker1972,RosnerTV1978}.
Understanding how magnetic fields will reconnect given a specific initial
configuration is therefore key to solar activity prediction. As observations
of the state of the solar magnetic field become increasingly well
resolved, from satellite observations such as SOHO and TRACE, and upcoming
missions such as Solar-B and STEREO, it is vital that models
which can make use of this information are developed and tested
as predictors of solar activity.
A goal of this paper is to advance this vital
understanding and show potential areas for future study.
In addition, a main goal is to study fundamental properties of 3D
reconnection, in particular the effect of helicity conservation
on reconnection dynamics.

We focus here on two types of twisted flux tube reconnection: the
slingshot and the tunnel  
\citep{DahlburgAN1997,Linton2001}. The slingshot is simply the 3D flux tube
analog of the classical 2D reconnection interaction in which two nearly
straight fieldlines collide and reconnect singly with each other.
A simulation of such 3D flux tube slingshot reconnection is shown in 
Figure \ref{fig:rr6q1}: here a pair of cylindrical flux tubes collide 
at right angles to each other, and their field reconnects once at
the collision site.  The reconnected fieldlines `slingshot' 
away from the collision site, releasing magnetic energy by 
becoming shorter, and form a new pair of cylindrical flux tubes by
Figure \ref{fig:rr6q1}(f). In contrast, as shown in Figure \ref{fig:rr6q10}, 
the tunnel is a uniquely 3D phenomenon.
This interaction, discovered by \citet{DahlburgAN1997}, 
starts off in Figure \ref{fig:rr6q10}(a) in a configuration much
like that of the slingshot, but rather than reconnecting just once, 
the flux tube fieldlines all reconnect twice, at two different
places. This allows the flux tubes to exchange a
small section of their volume, between the two reconnection points. Upon
exchanging these sections, they reform again by Figure \ref{fig:rr6q10}(f), 
but on the other side of each other from where they were initially. 
An analytical model developed by \citet{Linton2002}
showed that the tunnel occurs because helicity conservation
requires that the twist of the flux tubes decreases upon tunneling
in this configuration: the reduction of twist reduces the magnetic
energy, making the tunnel energetically favorable.

This paper studies the tunnel and slingshot interactions over 
a range of flux tube twists, collision angles, and resistive Lundquist 
numbers. To aid in this exploration, and to bring it into focus, we 
test the predictions which the analytical model of \citet{Linton2002} 
makes regarding these two reconnection interactions.
First we test the prediction that the slingshot should 
transition to the tunnel as twist is increased.  Second, 
we test the prediction that the tunnel should occur for flux tubes 
crossing at oblique angles as well as at right angles.
In parallel with this exploration, we measure the energy released
by this reconnection and compare it with the energy release predicted
by the analytical model. In \S 2 we introduce the simulations,
while in \S 3 and \S4 we introduce the slingshot and tunnel
interactions in detail. Then, in \S 5 we present the exploration
of reconnection versus twist for the orthogonal tube collisions,
as in Figures \ref{fig:rr6q1} and \ref{fig:rr6q10}. In \S 6 and \S 7 we 
repeat this exploration for two oblique collision angles at which our
theory predicts the tunnel should also occur, and we summarize our 
conclusions in \S 8.

\section{Simulations}

The flux tube simulations are performed with the CRUNCH3D code
\citep[see][]{Dahlburg1995}, on the Cray T3E at AHPCRC and ARSC, with a 
grant of computer time from the DoD.  This is a viscoresistive, compressible
MHD code. It is triply periodic and employs
a second-order Runge-Kutta temporal discretization and
a Fourier collocation spatial discretization. The simulations,
unless otherwise stated, are performed at a resolution
of $128^3$ modes.
The governing equations for this  compressible
MHD system are (as adapted from Dahlburg \& Norton 1995):
\bm
\begin{equation}\label{E:mass}
    \frac{\mbox{\ubm$\partial \rho$}}
      {\mbox{\ubm$\partial t$}}
       \mbox{\ubm$=-$}\nabla \cdot
       \mbox{\ubm$(\rho{\bf v}),$}
\end{equation}
\begin{equation}\label{E:momentum}
  \mbox{\ubm$\rho$}
  \frac{\mbox{\ubm$d{\bf v}$}}{\mbox{\ubm$dt$}}
  \mbox{\ubm$=$} \frac{\mbox{\ubm${\bf J\times B}$}}{\mbox{\ubm$c$}}
  \mbox{\ubm$-$}\nabla\mbox{\ubm$p +\rho\nu$}\nabla\cdot\tau\mbox{\ubm$,$}
\end{equation}
\ubm
\begin{equation}\label{E:magnetic}
  \frac{\partial{\bf A}}{\partial t}
    ={\bf v \times B}- \frac{4\pi\eta}{c}{\bf J},
\end{equation}
\bm
\begin{equation}\label{E:energy}
    \frac{\mbox{\ubm$\partial E$}}{\mbox{\ubm$\partial t$}}
       \mbox{\ubm$=-$}
     \nabla\cdot\left(
  \frac{\mbox{\ubm$\rho |{\bf v}|^2$}}{\mbox{\ubm$2$}}\mbox{\ubm${\bf v}+$}
  \frac{\mbox{\ubm$\gamma$}}{\mbox{\ubm$\gamma-1$}}
  \mbox{\ubm$p{\bf v}-\kappa$}\nabla\mbox{\ubm$ T
   -\nu\rho{\bf v}$}\cdot\tau
  \mbox{\ubm$+$}\frac{\mbox{\ubm${\bf B}$}}{\mbox{\ubm$4\pi$}}\times
  \left[\mbox{\ubm${\bf v\times B}$}
    \mbox{\ubm$-$}
   \frac{\mbox{\ubm$4\pi\eta$}}
      {\mbox{\ubm$c$}}\mbox{\ubm${\bf J}$}\right]\right)
   \mbox{\ubm$,$}
\end{equation}
\ubm
\begin{equation}
   E = \frac{\rho |{\bf v}|^2}{2}+\frac{p}{\gamma-1}
       + \frac{|{\bf B}|^2}{8\pi}.
\end{equation}
To preserve \bm$\nabla\cdot$\ubm${\bf B}=0$, the code evolves the vector
potential ${\bf A}$, with the magnetic field calculated
from ${\bf B} = $\bm$\nabla\times$\ubm$ {\bf A}$ whenever it is needed.
Here ${\bf v}$ is the flow velocity, $p$ is the
plasma pressure, $\rho$ is the density, $T$ 
is the temperature, $E$ is the energy density, 
${\bf J} = c$\bm$\nabla\times$\ubm$ {\bf B}/4\pi$ is the 
current, $\tau_{i,j}$ is the viscous stress tensor,
$\gamma=5/3$ is the adiabatic ratio, and $c$ is the speed of light.
Uniform thermal conductivity ($\kappa$) and kinematic viscosity
($\nu$) are assumed, but the magnetic resistivity ($\eta$) can 
have a spatial dependence.

The flux tubes we simulate here are constant twist, force free tubes,
also called Gold-Hoyle flux tubes \citep{Gold1960}. In coordinates 
centered on the axis of each flux tube, their magnetic field profile is
\begin{equation}\label{eq:axialfield}
    B_{axial}(r)=\frac{B_0}{1+\tw^2r^2}
\end{equation}
\begin{equation}\label{eq:azfield}
    B_{azimuthal}(r)=\tw rB_{axial}(r),
\end{equation}
for $0\leq r\leq R$, and 
the magnetic field outside the radius $r=R$ is set to zero.
The twist parameter ($\tw$) measures the winding of fieldlines
about the tube axis. Here lengths are normalized such that $\tw$
measures the number of times a fieldline
winds about the tube axis over $L=2\pi$ of axial distance,
where $L$ is the length of the box in all three dimensions.
We initialize the system with a uniform pressure
inside the flux tubes of $p_i=20/3$ in units
where the magnetic field strength on axis is $B_0/\sqrt{8\pi}=4$.
This gives a ratio of plasma to magnetic pressure on axis
of $\beta\equiv 8\pi p_i/B_0^2=.42$. 
To create pressure balance across the tube boundary, where
the magnetic field drops to zero, we initialize the external pressure, 
$p_e$, to be uniform with a value:
\begin{equation}
  p_e=p_i+\frac{B_0^2}{8\pi (1+\tw^2R^2)}.
\end{equation}
The ideal gas law $p=\rho \mathcal{R}T$ is assumed, where $\mathcal{R}$ 
is the ideal gas constant. The density is initialized as 
$\rho=\rho_0 p/p_i$, with $\rho_0=2$, so that the simulation volume 
is initially isothermal.  We choose the viscosity and magnetic resistivity
to be as low as possible while keeping the code stable.
The viscosity is $S_{\nu}\equiv v_AR/\nu=2880$, and
the resistive Lundquist number $S_{\eta}\equiv v_AR/\eta$
varies from $576$ to $288,000$. Here $v_A=B_0/\sqrt{4\pi\rho_0}$ 
is the Alfv\'en speed on axis, and we have 
chosen the flux tube radius as the typical system scale length.

Each simulation is initialized with a pair of flux tubes
with axes at $x=\pm\pi/4$, 
where the coordinates $[x,y,z]$ each range from $-\pi$ to $\pi$.
Each tube's radius is $R = 11\pi/48$, so 
the tubes' fields are initially separated by a distance $\pi/24=2R/11$. 
Flux tube 1, at $x=-\pi/4$, always has its axial field directed 
along the $-{\bf\hat z}$ direction. Flux tube 2, at 
$x=\pi/4$, has its axial field aligned at an angle $\theta$
to the axis of flux tube 1, where $\theta$ is measured in the left handed
sense about the ${\bf\hat x}$ axis, $i.e.$ in the clockwise
direction from the viewpoint of Figure \ref{fig:rr6q10}.  Each 
configuration is 
represented by the notation RRX where R denotes a right handed
flux tube, and $\theta=X\pi/4$. The right handed flux tube pairs shown 
in Figures \ref{fig:rr6q1} and \ref{fig:rr6q10} cross each other at an 
angle $\theta=6\pi/4$, and therefore are denoted as RR6. 

This initial equilibrium is perturbed with
a solenoidal velocity field, composed of two superimposed
stagnation point flows, which pushes the centers of
the two tubes toward each other at $x=y=z=0$:
\begin{equation}
   {\bf v}(x,y,z)=v_{0}[-\sin{x}(\cos{y}+\cos{z})~{\bf\hat x}
            +\cos{x}(\sin{y}~{\bf\hat y}+\sin{z}~{\bf\hat z})].
\end{equation}
This is not a driven flow, but rather is initialized at the
start of the simulation, with an amplitude $v_{0}=v_{A}/40$,
and then evolves dynamically as prescribed by the momentum
equation (eq. \ref{E:momentum}).

The helicity of the simulation is calculated via
\be\label{eq:helicity}
{\mathcal H}=\int {\bf A\cdot B} dV,
\ee
where the integral is taken
over the volume of the simulation. \citet{Hornig2005}
shows that the helicity can be calculated in this simple
manner for a triply periodic box on the condition that
magnetic flux penetrates at most two of the three sets of boundaries.
In that case, the periodic box can be mapped to a volume
enclosed within a pair of toroidal shells, as sketched in Figure \ref{fig:torusmap}, 
so that no flux leaves the closed volume. For example, if
in Figure \ref{fig:torusmap} flux penetrates the periodic 
side boundaries of the cube at $f1=f6$ and $f2=f5$ but not $f3=f4$, 
then $f1$ can be wrapped 
around to meet $f6$ and $f2$ can be wrapped around to $f5$, forming
the toroidal domain shown on the right. The result is that the flux penetrating these 
boundaries never leaves the toroidally mapped volume.  Thus, in such a 
mapping, the helicity is well defined. 
A necessary constraint on this formulation is that no flux thread 
the torus outside the simulation volume \citep[see][]{Hornig2005}. This
means that no flux can thread the volume poloidally through the `donut hole'
(through the area inside the grey ring),
and no flux can thread it toroidally along the innermost torus
(through the area inside the black ring).
To ensure this, the gauge for ${\bf A}$ is chosen so that the integrals 
of the vector potential along the lines around the outer edges of these areas,
{\it i.e.} along the intersection of boundary $f1$ with $f4$ and of boundary
$f2$ with $f3$, are zero:
\be
\oint_{f1\bigcap f4} {\bf A\cdot dl}=0=\oint_{f2\bigcap f3} {\bf A\cdot dl}.
\ee
A consequence of this choice of mapping and gauge is that 
when a pair of untwisted flux tubes cross as in Figure \ref{fig:rr6q1} with
a horizontal tube on the positive $x$ side of a vertical tube, the tubes 
do not link each other in the toroidal mapping, 
and so have a helicity of zero. 
If the $x$ positions of the tubes are switched, they
do link each other in the mapping, and so have a linking number
${\mathcal L}={\mathcal H}/\Phi^2=2$, where $\Phi$ is the
axial magnetic flux per flux tube. 
Additionally, a single untwisted flux tube crossing the domain diagonally 
in the ${\bf \hat{y}} + {\bf \hat{z}}$ direction will have a 
helicity of ${\mathcal H}/\Phi^2=-1$, 
whereas a flux tube crossing on the orthogonal diagonal will have
a helicity of ${\mathcal H}/\Phi^2=1$. Taking this into account,
we have verified that the simulation calculates the correct helicity
for each of the simulations reported here, and we can
track the evolution of the helicity during the flux tube interactions.
These values for the helicity can be different from what is
obtained from calculating the relative helicity with respect
to a potential reference field \citep[e.g.][]{BergerF1984, FinnA1985}, 
but the meaningful measures are the change in total, 
crossing, and twist helicity
from one state to another, and these are the same under both formalisms.
The first advantage of this toroidal mapping method is that it is
much simpler to implement. The second advantage is that it
is uniquely defined for triply periodic domains wherein a net magnetic
flux crosses only two of the three sets of boundaries, as is
the case for our simulations. In contrast, \citet{Berger1987} showed that the
relative helicity formulation of \citet{BergerF1984} and \citet{FinnA1985} 
is not in general uniquely defined for 
triply periodic domains unless no net magnetic flux crosses any of the 
boundaries.

\section{Slingshot Reconnection}

\begin{figure}[ht]
\epsscale{1.}
\plotone{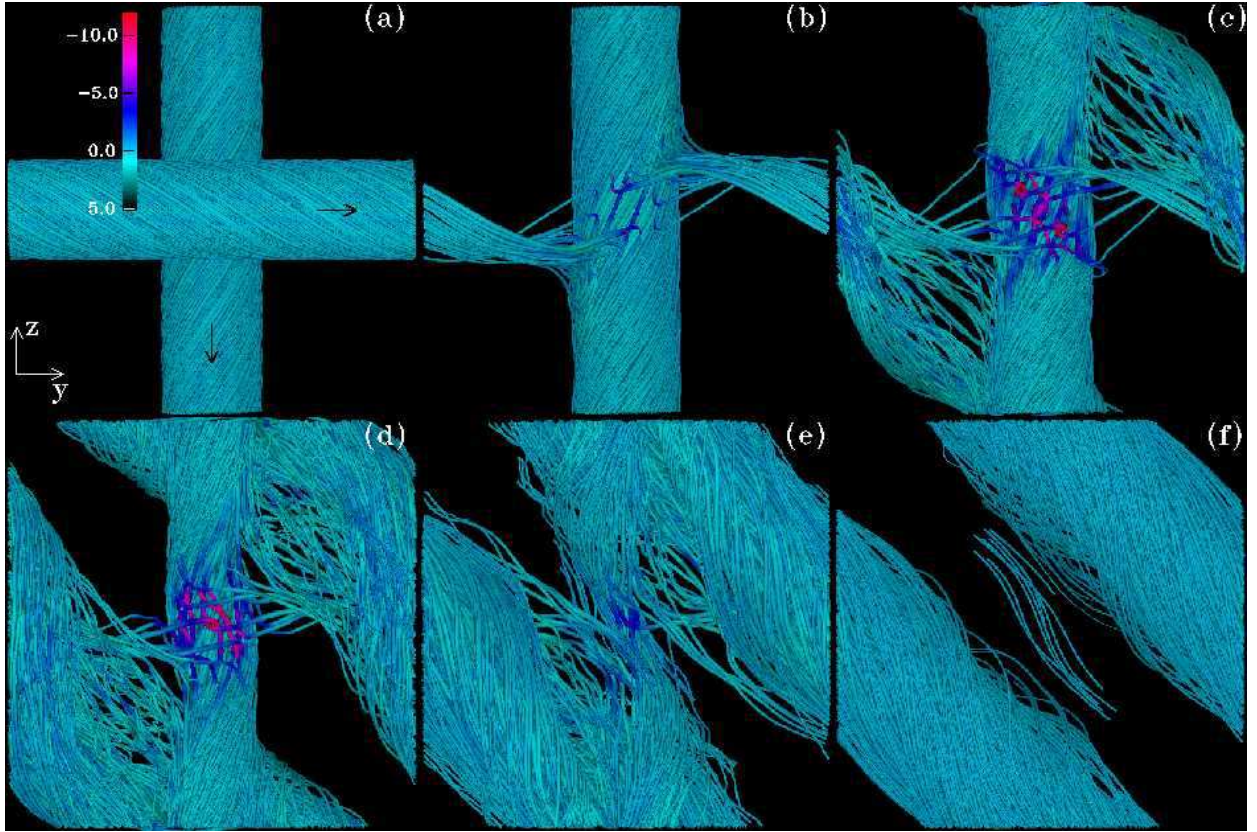}
\caption{Fieldlines from trace particles in a slingshot reconnection.
This is an RR6 collision at ${\mathcal T}=1$, with a uniform Lundquist 
number of $S_{\eta}=2880$. The color scale shows the parallel electric
current, which is strongest where reconnection is occurring. 
The interaction is shown at 
$tv_A/R=[0,12,26,38,57,115]$.
\label{fig:rr6q1}}
\end{figure}

We will now discuss the slingshot interaction shown in Figure \ref{fig:rr6q1} 
in more detail, and address an issue this simulation brings
up regarding the use of a periodic vector potential to calculate
the magnetic field. Figure \ref{fig:rr6q1} shows the interaction of  
RR6 tubes with a twist of $\tw=1$ at a uniform Lundquist number of
$S_{\eta}=2880$. Figure \ref{fig:rr6q1}(a) shows the initial configuration:
the fieldlines of both tubes are displayed, and one can
see that they are both right handed and wind about the tube
axes once over the length of the simulation box. These fieldlines
are traced from four $16^2$ grids of trace particles initially placed near both
ends of both flux tubes. As the simulation progresses, these
points move dynamically subject to the momentum equation 
(eq. [\ref{E:momentum}]).
To the extent that the fieldlines are frozen into the flow, this 
allows us to follow the dynamics of fieldlines.  Subsequently, in Figures 
\ref{fig:rr6q1}(b) through \ref{fig:rr6q1}(f), only fieldlines
traced from the trace particles initially on the vertical flux tube
are shown. This allows the reconnection region between the tubes
to be seen in the figure. In addition, it makes reconnected fieldlines
stand out: fieldlines connecting to the horizontal boundaries
are visible only if they have reconnected and are attached
to the vertical flux tube.  Figure \ref{fig:rr6q1}(b) shows the reconnection 
starting at the point where the two tubes first come into contact. 
The reconnected fieldlines form a sharp angle near 
the reconnection region, because the slingshot action of these 
fieldlines has not yet pulled them away from the reconnection region.  
By Figure \ref{fig:rr6q1}(c) the tension force has snapped 
these fieldlines away from the site of their reconnection,
converting magnetic energy into kinetic energy, which then
diffuses away via viscous drag.
This reconnection and snapping away continues through Figures 
\ref{fig:rr6q1}(d) and \ref{fig:rr6q1}(e) until, by Figure \ref{fig:rr6q1}(f), 
all of the field has reconnected to form a new pair of flux tubes.  
Throughout this figure, the parallel electric current 
$J_{\parallel}={\bf J\cdot B}/|{\bf B}|$ is shown by the color, 
with bright red being the strongest (negative) parallel
current as shown by the color bar. The values of parallel current
are normalized here by the initial current on axis $J_0=cB_0/(2\pi\tw)$. 
Parallel currents are a key signature of 3D reconnection 
\citep{Schindler1988},
and so this color highlights the regions of strong reconnection,
which are particularly evident in Figures \ref{fig:rr6q1}(c) 
and \ref{fig:rr6q1}(d). 

As the wavelength of the twist is significantly 
longer than the length of the collision interface, 
this reconnection should closely resemble the reconnection of 
untwisted flux tubes. Indeed, locally, the field
does reconnect in the straightforward fashion one expects
from untwisted reconnection. But there is a significant difference
here from the untwisted reconnection simulated by \citet{Linton2003}.
In the untwisted simulations, the flux tubes 
flatten out on contact and break up into several pieces
so that only a fraction of the flux reconnects. Here, in contrast, 
even this small amount of twist is strong enough to keep the tubes 
from flattening out and breaking up, so that all of the flux 
reconnects to form a new pair of isolated flux tubes.
Thus, the twist does not alter the local, relatively
small scale dynamics of the reconnection region, but it does act on the large
scale dynamics of the flux tubes to keep them coherent throughout
the reconnection process, and so changes the global reconnection
in a fundamental way.

Note that the reconnection seen here apparently contradicts the 
reconnection results reported by \citet{DahlburgAN1997}.  
They found that the collision of flux tubes in an apparently
identical configuration, RR6 with $\tw=1$ and the same Lundquist number
we use here, results in little or no reconnection: instead, the tubes
bounce off each other.  The resolution of this contradiction is that 
the initial conditions for their simulation have a small amount of 
background field, which interferes with the reconnection at small twist. 
This background field arises naturally
from the periodic boundary conditions on the magnetic vector
potential {\bf A}. Such a periodic vector potential can represent
a periodic magnetic field ${\bf B} = $\bm$\nabla\times$\ubm$ {\bf A}$, 
but it allows no net magnetic flux in any direction.
Formally, the net flux in, for example, the ${\bf\hat z}$ 
direction is the integral of $\bf{A}$ on the curve running 
around the edge of the simulation box at any value of $z=z_c=const$:
\be
\Phi_z=
\oint_{z_c}{ \bf A\cdot dl}=
\int_{-\pi}^{\pi} A_x|_{y=-\pi}dx
+\int_{-\pi}^{\pi} A_y|_{x=\pi}dy
+\int_{\pi}^{-\pi} A_x|_{y=\pi}dx
+\int_{\pi}^{-\pi} A_y|_{x=-\pi}dy.
\ee
Since the periodic boundary conditions mean that
$A_y(-\pi,y,z_c)=A_y(\pi,y,z_c)$ and 
$A_x(x,-\pi,z_c) =A_x(x,\pi,z_c)$, the two $dy$ integrals 
cancel each other, as do the two $dx$ integrals, and so there must be
no net flux in the ${\bf\hat z}$ direction. Instead, the magnetic
field which results from a periodic vector potential is
the specified field plus a uniform background
field whose net flux exactly cancels the net flux of the 
specified field. To correct for this 
background field effect, we calculate ${\bf B}$ at each step as
\bm
\be
     \mbox{\ubm${\bf B} =$} \nabla\times 
      \mbox{\ubm${\bf A} + B_{0x}{\bf \hat{x}}
 +B_{0y}{\bf \hat{y}}+B_{0z}{\bf \hat{z}}$}
\ee
\ubm
where the three background field constants are calculated
at the start of the simulation, when the ${\bf B}$ field
is specified, as
\be
B_{0i}=\int_{-\pi}^{\pi}\int_{-\pi}^{\pi}\frac{{\bf B}|_{x_i=const}}{4\pi^2}
          {\bf\cdot\hat{x}_i}dx_jdx_k.
\ee  
These corrections are then saved for use during the rest of 
the simulation, and {\bf A} is calculated from the inverse 
laplace transform of {\bf B}:
\bm
\be
\nabla \times \mbox{\ubm${\bf B} = $}
\nabla \times \nabla \times \mbox{\ubm${\bf A} = \nabla^2 {\bf A},$}
\ee
\be
 \mbox{\ubm${\bf A} = $}  \nabla \times 
       \frac{\mbox{\ubm$1$}}{\mbox{\ubm$\nabla^2$}} \mbox{\ubm${\bf B}.$}
\ee
\ubm
From then on, ${\bf A}$ is evolved dynamically by the code,
ensuring that the solenoidality of ${\bf B}$ is preserved.

An obvious question that this formalism raises is:
does the lack of any presence of the background field 
in the periodic vector potential affect the dynamics? 
Only the induction equation could be affected, as
the other dynamical equations rely on ${\bf B}$ to calculate
magnetic effects rather than ${\bf A}$. Fortunately, the induction equation,
\begin{equation}
  \frac{\partial{\bf A}}{\partial t}
    ={\bf v \times B}- \frac{4\pi\eta}{c}{\bf J},
\end{equation}
can also be shown to be unaffected by the absent linear component of ${\bf A}$.
Both the source terms ${\bf J}$ and ${\bf v\times B}$ make no linear 
contribution to this equation, as ${\bf J}$, ${\bf v}$, and ${\bf B}$ 
are all periodic in space, 
and so have no linear components. Thus the linear part of the induction
equation is reduced to $\partial{\bf A}_{linear}/\partial t = 0$, which
just confirms our assumption that ${\bf A}_{linear}$ and therefore
the background magnetic correction terms $B_{0i}$ are constant in time.
Note that the helicity calculation of equation (\ref{eq:helicity})
is the one place where the linear corrections to the vector potential
are needed: for this integral the correction
${\bf A}_{linear}={\bf\hat y}B_{0z}x-{\bf\hat z}B_{0y}x$ 
are therefore added to ${\bf A}$ (assuming that $B_{0x}=0)$.

We have carefully checked the resulting magnetic field,
and found that it matches our input conditions.
We have also checked the field without this correction
and found that there is indeed zero net flux in all directions.
The correction has been included in the simulations reported here,
in \citet{Linton1998, Linton1999, Linton2001}, and in \citet{Linton2003}, but
not in the simulations performed by CRUNCH3D before that time.
Thus for the \citet{DahlburgAN1997} simulations, with flux
tubes directed in the $-{\bf \hat z}$ direction and
in the ${\bf \hat y}$ direction, there is an extra background field
directed diagonally across the whole simulation volume in the 
$-{\bf\hat y}+{\bf\hat z}$ direction.  The ${\bf \hat z}$ 
background component, for example, is
\be\label{eq:b-background}
B_{z0}\equiv\frac{\int B_z dxdy}{\int dxdy}=
\frac{\pi B_0 \ln(1+\tw^2R^2)}{\tw^2 L^2}.
\ee
For $\tw=1$ this gives $B_{z0}=.13$ versus a peak axial field strength in
the flux tube of $B_0=4$. 
A part of this field lies between the tubes and so provides a force 
to block them from coming in contact. We have
performed this simulation with this correction turned off,
and replicated the flux tube bounce seen by \citet{DahlburgAN1997}.  
We then turned the correction
on and found the slingshot interaction shown in Figure \ref{fig:rr6q1} 
and discussed above.  Note that for $\tw=10$, equation (\ref{eq:b-background})
this gives $B_{z0}=.013$, so the effect is weaker in that 
case than in the low twist case, likely explaining why the flux
tubes reconnected at $\tw=10$ to tunnel in \citet{DahlburgAN1997}
in spite of the background field.

\section{Tunnel Reconnection}

\begin{figure}[ht]
\epsscale{1.}
\plotone{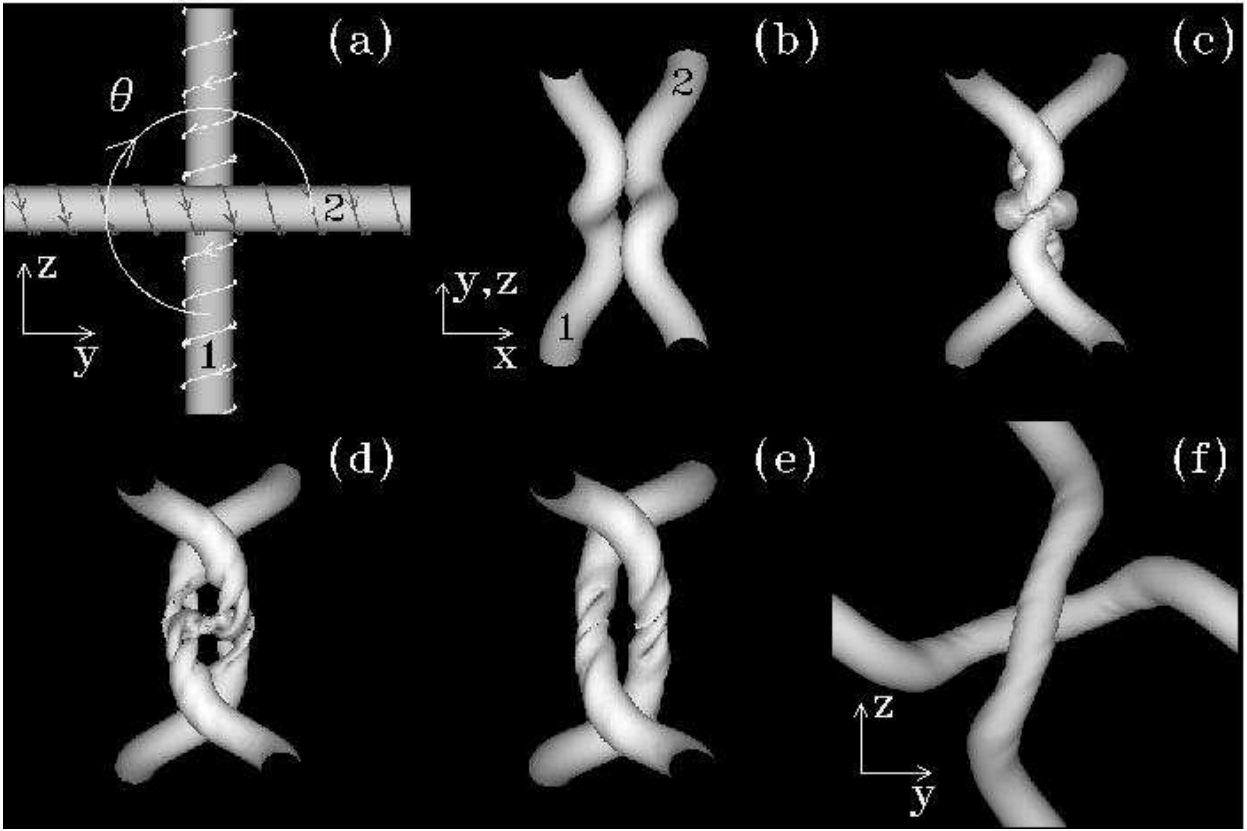}
\caption{Isosurfaces at $|{\bf B}|_{max}/2$ from a tunnel
reconnection. This is from an RR6 simulation at 
${\mathcal T}=10$, at a uniform Lundquist number of $S_{\eta}=2880$. 
Flux tube 1 is parallel to the $-{\bf\hat{z}}$ axis and flux
tube 2 is at an angle $\theta=6\pi/4$ relative to flux tube 1.
This shows how the tubes reconnect twice to exchange center sections
and tunnel.  Note that the viewpoint changes from panel (a) to panel (b),
and then changes back again in panel (f).
The times shown are $tv_A/R=[0,25,42,52,61,70]$.
\label{fig:rr6q10}}
\end{figure}

\begin{figure}[ht]
\epsscale{1.}
\plottwo{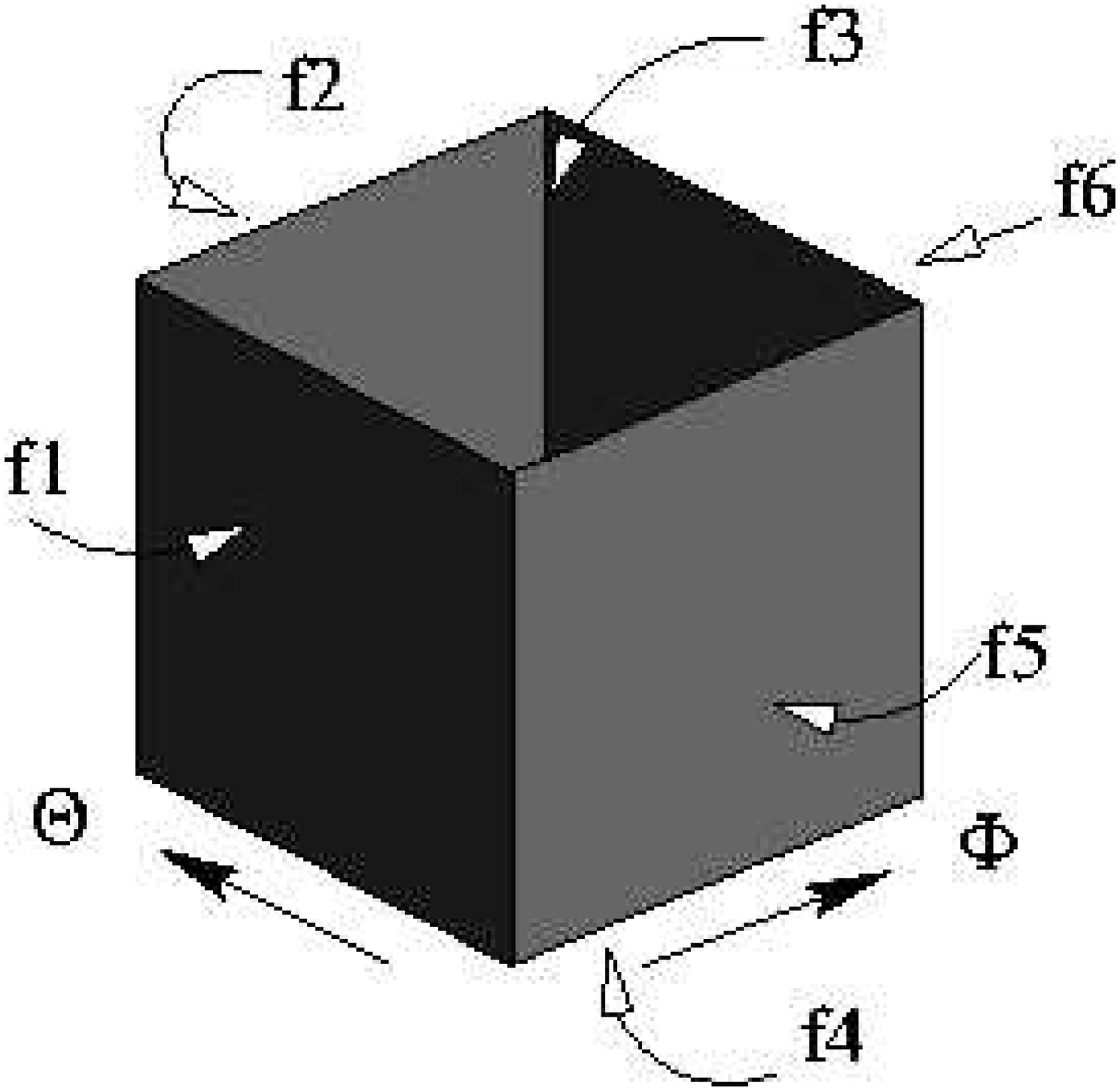}{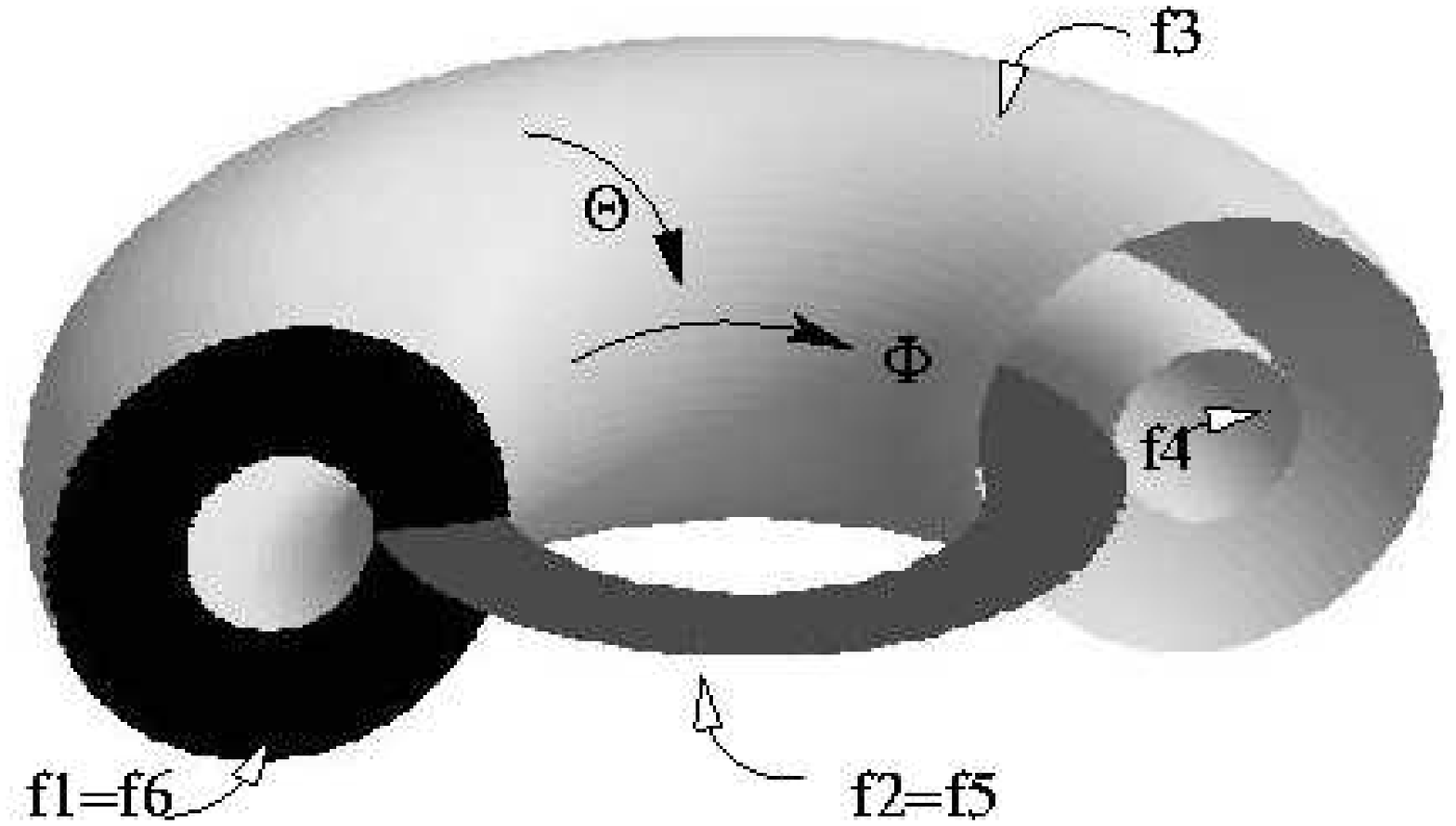}
\caption{Sketch of the toroidal mapping of a doubly periodic 
volume. In the mapping, surface $f1$ is joined to surface $f6$ 
while surface $f2$ is joined to surface $f5$. If no flux
penetrates the boundaries of the toroidal volume, then the
helicity is simply $ {\bf A\cdot B}$ integrated over the volume.
\label{fig:torusmap}}
\end{figure}

\begin{figure}[ht]
\epsscale{1.}
\plotone{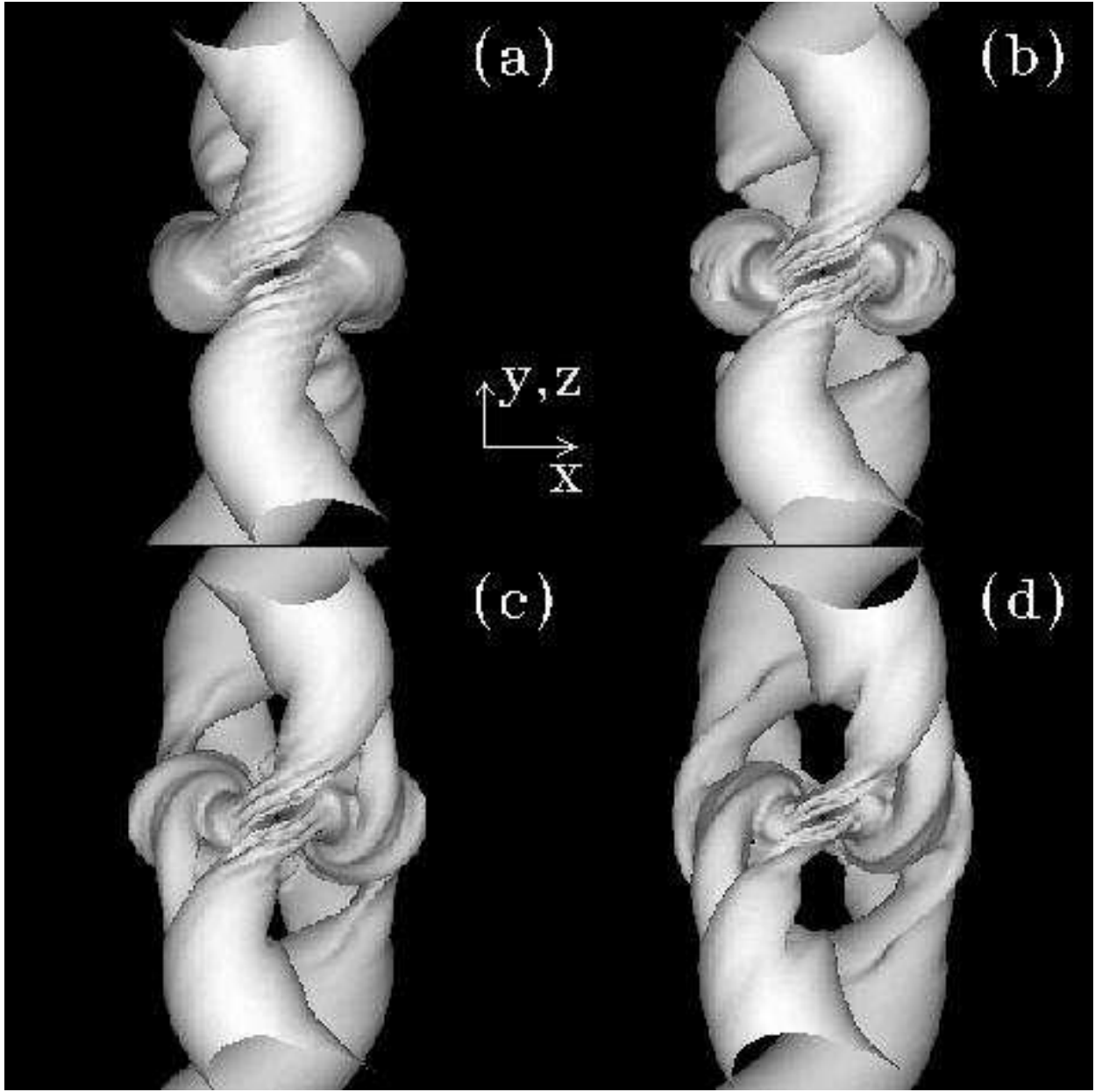}
\caption{Closeup view of Figure \ref{fig:rr6q10}
 at $tv_A\pi/R=[42,47,49,52]$.
\label{fig:rr6q10loc}}
\end{figure}

Having introduced the slingshot reconnection, we will now introduce
the tunnel reconnection, and discuss the theory of why it occurs.
Figure \ref{fig:rr6q10} shows isosurfaces at $|{\bf B}|=|{\bf B}|_{max}/2$ for
a simulation of an RR6 collision at $\tw=10$, a uniform resistivity of 
$S_{\eta}=2880$, and at a resolution of $256^3$. 
Figures \ref{fig:rr6q10}(a) and \ref{fig:rr6q10}(f) show the 
beginning and end of the simulation from the same viewpoint
(from the ${\bf\hat x}$ axis). These panels show that,
as seen by \citet{DahlburgAN1997}, the flux tubes reconnect 
such that they tunnel entirely through each other.  To make the reconnection 
more visible, the intermediate panels show the interaction as seen from the 
upper left hand corner of the view shown in Figure \ref{fig:rr6q10}(a).
In Figure \ref{fig:rr6q10}(b) the tubes have just collided, and are rebounding,
exciting the helical kink instability. Figure \ref{fig:rr6q10}(c) shows
that, before they can completely rebound, the tubes start to reconnect at
two places. To show the reconnection in more detail, the center of Figure 
\ref{fig:rr6q10}(c) is shown in a closeup view in Figure \ref{fig:rr6q10loc}(a), 
along with a closeup of Figure \ref{fig:rr6q10}(d) in Figure \ref{fig:rr6q10loc}(d) 
and a closeup of two intervening time steps in Figures \ref{fig:rr6q10loc}(b) and 
\ref{fig:rr6q10loc}(c). The two tubes reconnect on
the near side and the far side of this view. The bulging sections
of flux tube (to the left and right of the center of the collision) between 
the reconnection points do not appear to reconnect: they remain in evidence 
throughout Figures \ref{fig:rr6q10loc}(a)-\ref{fig:rr6q10loc}(d). However, as 
the reconnection proceeds, the connections to these sections change dramatically. 
In Figures \ref{fig:rr6q10loc}(a) and \ref{fig:rr6q10}(c), the near flux tube 
at the bottom is attached to the bulging section on the right, and then to 
the far flux tube at the top of the simulation.  But by Figures 
\ref{fig:rr6q10loc}(d) and \ref{fig:rr6q10}(d), the near flux tube at the bottom 
is attached to the {\it left} bulging section and then, as before, to the 
far flux tube at the top. So the top and bottom parts of this flux tube
are the same as at the beginning, but the middle section
has been exchanged with that of the other flux tube, resulting
in a change of position of the two flux tubes. In this manner,
the flux tubes appear to tunnel through each other.
Note that the inverse 
reaction, where tubes in the configuration of 2(f) collide with
each other, does not result in a tunnel.
\citet{Linton2001} simulated this RR2 collision and found that, 
at a twist of $\tw=10$, the flux tubes simply bounce off each other. 
This tunnel interaction therefore raises two questions: 1) why
does it occur and 2) why is it not reversible?

To address these questions, \citet{Linton2002} explored the 
tunneling interaction analytically, assuming helicity conservation, and 
using the concepts of magnetic twist and linking or crossing number developed 
by \citet{BergerF1984}. They found a promising explanation: by 
tunneling, flux tubes change how they link
each other, as measured by the linking number ${\mathcal L}$ 
\citep[see][]{WrightB1989}.  
As mentioned earlier, for the helicity formalism we use here, the flux 
tubes in Figure \ref{fig:rr6q10}(a) have ${\mathcal L}=0$, wheres those in Figure 
\ref{fig:rr6q10}(f) have ${\mathcal L}=2$.  Thus in tunneling, their linking 
number increases by $2$. Note that in any another helicity formalism, the 
value of ${\mathcal L}$ may be different, but the change in ${\mathcal L}$ due 
to reconnection, which is the only meaningful number, will be exactly the same as it 
is in this formalism. The sum of the tubes' linking and twist numbers
is their total helicity, ${\mathcal H}/\Phi^2=2\tw+{\mathcal L}$, and
this sum must be conserved to conserve helicity. When the tubes 
tunnel to increase their linking helicity, their twist helicity must decrease. 
For an RR6 configuration tunneling to RR2, the positive twist $\tw$ 
is reduced by one turn per flux tube upon tunneling, and 
the twist energy decreases as $\tw^2$.  In the opposite 
interaction, when RR2 flux tubes tunnel to an RR6 configuration, the 
linking number decreases by $2$, and so the twist must increase by
one turn per tube, increasing the magnetic energy. This inverse reaction is 
energetically unfavorable, and therefore not expected 
to occur, in agreement with the simulation results of \citet{Linton2001}.

\citet{Linton2002} found that the magnetic energy, normalized by
$4\tw^2/(L B_0^2)$, of the pair of Gold-Hoyle flux tubes we simulate
here is initially 
\be
 M_0=(1+\zeta)(\Gamma_{12}+\Gamma_{32})
\ee
where $L\zeta$ is the length of flux tube 2 in the 
simulation box.  When this tube crosses the box perpendicular to 
the boundaries, as in RR6, its length is $L$, the same
as the length of flux tube 1, and so $\zeta=1$.  
When this tube crosses the box at a diagonal, as in RR5 or RR7, its 
length is $\sqrt{2} L$ and $\zeta=\sqrt{2}$. 
The constants $\Gamma_{12}$ and $\Gamma_{32}$ come from integration of the 
square of the initial axial (eq. [\ref{eq:axialfield}]) and azimuthal 
(eq. [\ref{eq:azfield}]) magnetic field, respectively, over the tube cross
section, and have the form
\be
\Gamma_{12}=\frac{\tw^2R^2}{2+2\tw^2R^2}
\ee
and
\be
\Gamma_{32}=\frac{1}{2}
\left(\ln(1+\tw^2R^2)-\frac{\tw^2R^2}{2+2\tw^2R^2}\right).
\ee
From helicity conservation along with mass and
flux conservation, flux tubes initially crossing at angles in 
the range $\pi < \theta < 2\pi$ should have a normalized 
magnetic energy after tunneling of
\be\label{eq:tunnel}
 M_t=(1+\zeta)\Gamma_{12}+ \left( 
    \frac{(\tw\zeta-1)^2}{\tw^2\zeta}+\frac{(\tw-1)^2}{\tw^2} 
         \right)\Gamma_{32}.
\ee
Here the factors $\tw\zeta-1$ and $\tw-1$ take account of the loss of $1$
turn of twist in each flux tube due to tunneling.
Note that this result relies on the assumption that the tubes evolve
homologously: $i.e.$ that are they still uniform twist tubes
after reconnection. 
This equation shows that the tunnel changes the azimuthal energy ($\propto \Gamma_{32}$)
but not the axial energy ($\propto \Gamma_{12}$).
\citet{Linton2002} applied the same analysis to the slingshot 
tubes: in that case, the tubes lose only $1/2$ turn
of twist as they reconnect. However, the axial field can also lose
energy because the slingshot allows the tubes to shorten, as 
shown in Figure \ref{fig:rr6q1}.  They find that these tubes should 
have a normalized magnetic energy after slingshot of 
\be\label{eq:slingshot}
 M_s=\frac{4}{\zeta+1}\frac{L_s^2}{L^2}\Gamma_{12}+
    2\frac{L}{L_s}\frac{1}{\tw^2}
   \left(\tw\frac{\zeta+1}{2}-\frac{1}{2}\right)^2\Gamma_{32},
\ee
where $L_s$ is the length of the tube after slingshot. 
Here the factor $\tw(\zeta+1)/2-1/2$ accounts 
for the loss of $1/2$ turn of twist per flux tube in a slingshot.  

If we assume that flux tubes will reconnect in the manner which
releases the most energy, we can now predict whether a given
colliding flux tube pair will slingshot or tunnel when they
reconnect.  For this analysis, we use the optimal slingshot length $L_s$ in
calculating the slingshot energy, as in \citet{Linton2002}
equation (40). This length is calculated to be that which gives the lowest 
slingshot energy, given the geometrical constraint that the length cannot 
be shorter than the distance between the tube's footpoints. 
This optimal length is not always the shortest because,
as shown by equation (\ref{eq:slingshot}), the axial field energy of
the tubes (the $\Gamma_{12}$ term) varies as $L_s^2$ while
the azimuthal energy (the $\Gamma_{23}$ term) varies as $1/L_s$.
The optimal length is therefore a compromise between axial and
azimuthal (twist) energy.  

We plot the predicted energies from this analysis
as a function of twist for these two reconnected states, normalized
by the initial state, in Figure \ref{fig:rr6energy} for the RR6 collision. 
The resulting slingshot and tunnel energy 
curves are shown by the dashed-triple-dotted and by the dashed curves,
respectively. The tunneled state is at a lower energy for
high twists, because the twist energy dominates in that regime,
and the tunnel releases more twist energy than the slingshot.
The slingshot state, in contrast, is at lower energy for low twist,
as the axial field energy dominates there, and the slingshot
can release axial energy by making the tubes shorter, whereas the
tunnel cannot.  These two curves cross at $\tw=2.56$, predicting that 
at twists below this level the slingshot loses more energy than 
the tunnel.  If energy release alone dictates the resulting state, the 
interaction should be a slingshot reconnection below that twist and 
a tunnel above that twist. Obviously the dynamics between the beginning
and end state will play an important role, but
the expected energy release given by equations (\ref{eq:tunnel}) and 
(\ref{eq:slingshot}) should provide a useful predictor of the end state.

Do the simulations agree with these predictions?
In Figure \ref{fig:rr6q1} we show that the slingshot
occurs at $\tw=1$ and $S_{\eta}=2880$, and in Figure \ref{fig:rr6q10} 
we show that the tunnel occurs at $\tw=10$ and
$S_{\eta}=2880$, so, in agreement with our theory, there is clearly a 
transition at some level of twist between $1$ and $10$.  
However, \citet{DahlburgAN1997}
also found that the type of reconnection which occurs
depends on the Lundquist number: for $\tw=10$ RR6 flux
tubes, they found that at $S_{\eta}=2880$ a tunnel
occurs, but at $S_{\eta}=576$ a slingshot occurs. 
Clearly there is a dependence on resistivity as well as
twist. In the following section, we will present a series
of MHD simulations testing these analytical
predictions for the transition as a function of twist, and
will also explore the dependence of the transition on
resistivity.

\section{RR6 Reconnection}
For our first set of RR6 simulations, we searched for the transition 
from tunnel to slingshot reconnection as a function of twist at four 
different spatially uniform Lundquist numbers. At the lowest number 
we tested, $S_{\eta}=576$, we found only slingshot interactions even at 
twists as high as $\tw=14$.  This agrees with the finding by \citet{DahlburgAN1997} 
that a $\tw=10$ collision at this Lundquist number 
results in a slingshot. It also suggests that, since the tubes slingshot
even at very high twist, the tunnel will never occur for any twist
at such a low Lundquist number. However, by increasing the Lundquist 
number to $1440$, we did find that the tubes tunnel at twists equal to 
or greater than $6.5$.  Below this critical value, at twists equal to or 
below $6$, the slingshot occurs.  Note that our resolution in twist is only 
$\delta \tw=.5$, as an entire flux tube collision simulation is necessary 
for each data point, so high resolution in $\tw$ space is expensive. Clearly 
at moderate Lundquist numbers both the tunnel and slingshot can occur, 
though the transition from one to the other occurs at a significantly 
higher level than the prediction of $\tw=2.56$ from the analytical model.
As we increased the Lundquist number from $1440$, the critical
twist for the transition decreased: the tunnel occurs
down to a twist of $6$ at a Lundquist number of $2880$
and to a twist of $5.5$ at a Lundquist number of $5760$.
We postulated that this decrease in critical twist as a function
of Lundquist number is because the twist diffuses
away due to the resistivity: the lower the Lundquist number,
the faster the twist disappears during the simulation. In effect,
enough twist may disappear before the tubes have a chance to tunnel that 
the tubes see a much lower level of twist than the code was initialized 
with.  In fact, for the $\tw=5.5$ tunnel simulation at $S_{\eta}=5760$ the 
helicity decreases to $53\%$ of its initial value by the time the tubes 
finish tunneling. As the linking number is zero in this configuration, 
the helicity is purely due to the twist, and so this indicates that the 
twist per tube diffuses from $5.5$ to $2.9$ during that time.
At the very low Lundquist number of $576$, twist would diffuse
away at an even faster rate. Thus it is not surprising that the
tubes do not tunnel at such a low Lundquist number even if their initial
twist is very high.

As the code becomes unstable
at Lundquist numbers higher than $5760$ for these simulations,
we could not decrease the resistivity further to see if
the transition eventually goes to the predicted level 
of $\tw=2.56$ as the Lundquist number gets very large. Instead we turned to the
possibility of a nonuniform resistivity. We modified the code so that 
it has a ball of high resistivity at the center of the simulation box, where 
the flux tubes collide, of the form
\be
\eta=\eta_0(1+99e^{-\lambda^2/(4R^2)}),
\ee
where $\lambda$ is the radial distance from the center of the box: 
$\lambda=(x^2+y^2+z^2)^{1/2}$.
The diameter ($2R$) of the flux tubes is used as the scale length so
that the collision area of the tubes will mostly be within
the high diffusion area. This ensures that the region
of intense dynamics has sufficient diffusion that 
the reconnection does not destabilize the code,
but allows the majority of the flux tubes' volume, which is
well away from the reconnection, to diffuse at a much slower rate.
This worked very well, and we were able to run the code
at a background Lundquist number of $288,000$, with a minimum
Lundquist number in the collision region of $2880$. In
this case, tubes at $\tw=5.5$ tunnel in about $1/5$ the time it takes
them to tunnel at a uniform resistivity of $S_{\eta}=5760$,
and the helicity only decreases by $6\%$.  This indicates two important results. 
First, the loss of helicity is clearly due to diffusion rather than to reconnection, 
as both the high and low Lundquist number simulations undergo the same tunnel
reconnection, but the low Lundquist number simulation loses about $8$ times
more helicity than the high Lundquist number simulation.
Second, at a loss of only $6\%$, the twist is reduced from $5.5$ to
$5.2$ due to diffusion. This difference is well below our resolution
of $\delta\tw=.5$, and so diffusive loss of twist should have a negligible
effect on the tunnel to slingshot transition.
We still found, however, that the tubes tunnel only down to a twist 
of $5.5$ and then slingshot for twists of $5$ and below. The 
true simulation limit at which the tunnel can occur for an RR6 collision
therefore appears to be in the range $\tw=5$ to $5.5$, 
as opposed to the analytical prediction of $2.56$.

Figures \ref{fig:rr6q5} and \ref{fig:rr6q5.5} show the simulations on either 
side of this transition from tunnel to slingshot for these nonuniform 
resistivity simulations with $S_{\eta}=288,000$ outside the collision region. 
Figure \ref{fig:rr6q5} shows the $\tw=5$ simulation: as before, only the 
fieldlines from the trace particles in the vertical tube are plotted. 
The inset in Figure \ref{fig:rr6q5}(a) shows the initial tube isosurfaces at 
$|{\bf B}|=|{\bf B}|_{max}/3$, with arrows superimposed on the tubes to show 
the axial field directions.  The field from the tubes reconnects in Figures
\ref{fig:rr6q5}(b) through \ref{fig:rr6q5}(d), and  slingshots in Figures 
\ref{fig:rr6q5}(e) and \ref{fig:rr6q5}(f) to form a pair of diagonal, reconnected
tubes. An isosurface of the field in the final state is inset into Figure
\ref{fig:rr6q5}(f) for comparison with the initial state in 
Figure \ref{fig:rr6q5}(a): clearly the tubes have slingshot to a shorter length, 
though the twist remaining after reconnection is strong enough to kink the tubes 
into a helical shape.  The majority of the fieldlines simply slingshot because they
only reconnect once, but Figures \ref{fig:rr6q5}(e) and \ref{fig:rr6q5}(f) 
show a small number of fieldlines which have reconnected twice to tunnel. These
fieldlines connect the slingshotted tubes at their
centers and exert a tension force pulling the reconnected tubes back together. 
However this force is insufficient to bring the tubes close enough that 
they come in contact again: by Figure \ref{fig:rr6q5}(f) 
the configuration has settled into a static equilibrium, and the interaction
is over. Presumably, if the tubes had been pulled back into contact, their fieldlines 
would have reconnected again to tunnel to the lower energy state. This shows the 
limitations of the energy analysis: if the barrier to tunnel reconnection is
too large to overcome, as for example, when the tubes do not remain in contact
long enough for the fieldlines to reconnect twice, the tubes will not tunnel.

Figure \ref{fig:rr6q5.5} shows the $\tw=5.5$ RR6 flux tube interaction.
Most of the interaction is very similar to the $\tw=5$
interaction, as one would expect, considering how
small a difference in twist there is between the two
simulations. The tubes reconnect in Figures \ref{fig:rr6q5.5}(b)
through \ref{fig:rr6q5.5}(d) and start to slingshot in Figure 
\ref{fig:rr6q5.5}(e).  However, more of the fieldlines have 
reconnected twice to tunnel by this point than in the $\tw =5$ simulation. 
When the tension force of these fieldlines pulls the tubes back toward 
each other, they actually bring the tubes into contact again. The tubes can 
therefore reconnect a second time to tunnel by Figure \ref{fig:rr6q5.5}(f). 
The majority of the horizontal fieldlines are no longer plotted by 
Figure \ref{fig:rr6q5.5}(f) because they are no longer connected to the 
vertical tube as they were in the preceding panels: the horizontal tube
has been recreated with little or no connection to the vertical tube. 
The isosurface inset into Figure \ref{fig:rr6q5.5}(f) clearly shows that 
the field of this recreated horizontal flux tube is now behind the field of 
the vertical flux tube, in contrast to the initial state of Figure 
\ref{fig:rr6q5.5}(a), where it was in front. The fact that these tubes 
attempt to slingshot in Figure \ref{fig:rr6q5.5}(e) before recombining to 
tunnel confirms our conjecture in the previous paragraph: the slingshot flux 
tubes must be pulled back into contact again after their first reconnection if
they are to reconnect a second time and tunnel to the lowest energy
state. 

Figure \ref{fig:rr6energy2} shows the energetics of these two interactions.
From top to bottom, we plot the helicity (dash-triple-dot lines), the magnetic 
energy (solid lines), the kinetic energy (dashed and dash-dot lines),
and the time derivative of helicity (solid lines). In each case, the slingshot
and tunnel lines are labeled with `s' and `t', respectively.
A large peak in the kinetic energy at about $tv_A/R=40$, accompanied
by a significant drop in the magnetic energy shows that the slingshot 
reconnection is converting magnetic into kinetic energy in both simulations 
at this time.  After this point, the $\tw=5$ 
tubes, which only slingshot, continue to convert magnetic into kinetic energy 
and reach an even higher kinetic energy peak at just after $tv_A/R=50$.
Once the reconnection is over, the kinetic energy starts to drop.
This is partly due to viscous losses, but also due to harmonic
oscillation of the flux tubes: they overshoot their new optimal equilibrium
positions, and slow to a stop as the magnetic fields are overstretched.
Thus this drop in kinetic energy at $tv_A/R=80$ to $90$ is accompanied by
a brief rise in the magnetic energy. The tubes then rebound, briefly 
increasing their kinetic energy again before they settle into an equilibrium.
During this period, 
the magnetic energy of the tunneling tubes at $\tw=5.5$ also steadily 
decreases due to diffusion until the tubes come into contact again at
about $tv_A/R=110$. Then the reconnection starts again as the tubes tunnel: 
the magnetic energy decreases at a faster rate than it did during the 
preceding diffusion stage, and the kinetic energy shows another local peak.
This peak is, however, significantly lower than the peak from the slingshot 
reconnection in either simulation, as the tunnel is much less dynamically active 
than the slingshot. 

The helicity decreases very slowly during the whole simulation, but the
time derivative of helicity shows that the helicity decay rate
increases linearly for the first $30$ Alfv\'en times. This is
likely due to the increase in magnetic gradients between the
flux tubes during this interval, as the two tubes collide
and generate a current sheet between them. Next, from $tv_A/R=30$ 
to $tv_A/R=40$, the helicity decay rate slows rapidly, and then it
settles down to a nearly constant rate for the rest 
of the simulation. The slowdown is likely due to the destruction by 
reconnection of the sharp magnetic gradients induced by the initial collision.
For the remainder of the simulation, the slingshot helicity decays
at a slower rate than the tunnel helicity: this is likely because the
slingshot removes the tubes entirely from the central, high diffusion
region while the tunnel keeps the tubes near that region for the entire
interaction, so the tunnel tubes see a higher average resistivity and
decay at a faster rate.

\begin{figure}[ht]
\epsscale{1.}
\plotone{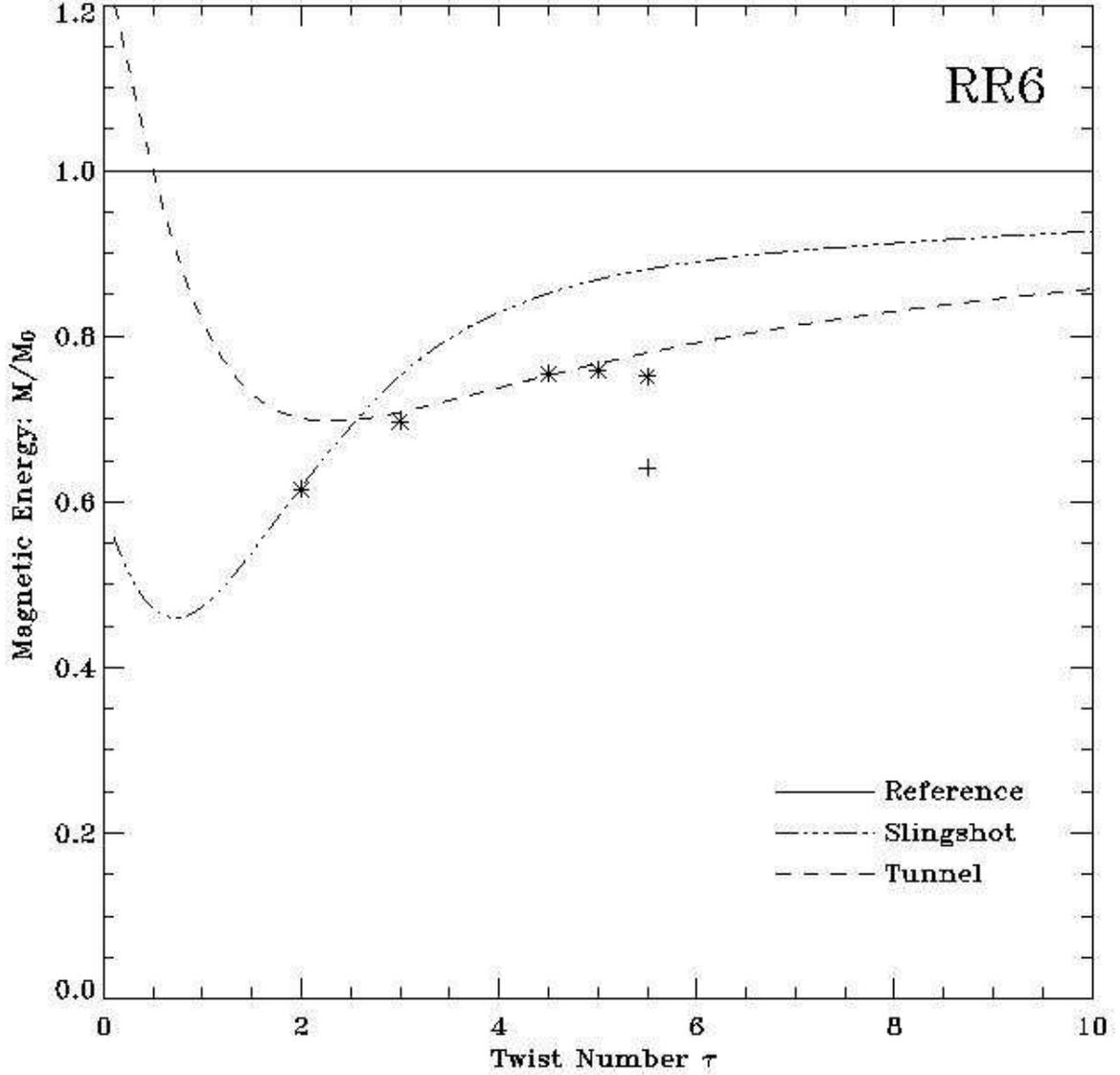}
\caption{Theoretical reconnection energy for RR6 simulations at various
twists. The dashed line shows the predicted tunnel energy, as 
calculated from equation (\ref{eq:tunnel}), and the dash-triple-dot
line shows the slingshot energy, as calculated from equation (\ref{eq:slingshot}). 
The asterisks denote simulation values for slingshot interactions, 
while the plus sign denotes the simulation value for the tunnel
interaction of Figure \ref{fig:rr6q5.5}. All the simulation data
shown here is from the nonuniform resistivity calculations, with
a background, maximum Lundquist number of $L_{\eta}=288,000$.
\label{fig:rr6energy}}
\end{figure}

\begin{figure}[ht]
\epsscale{1.}
\plotone{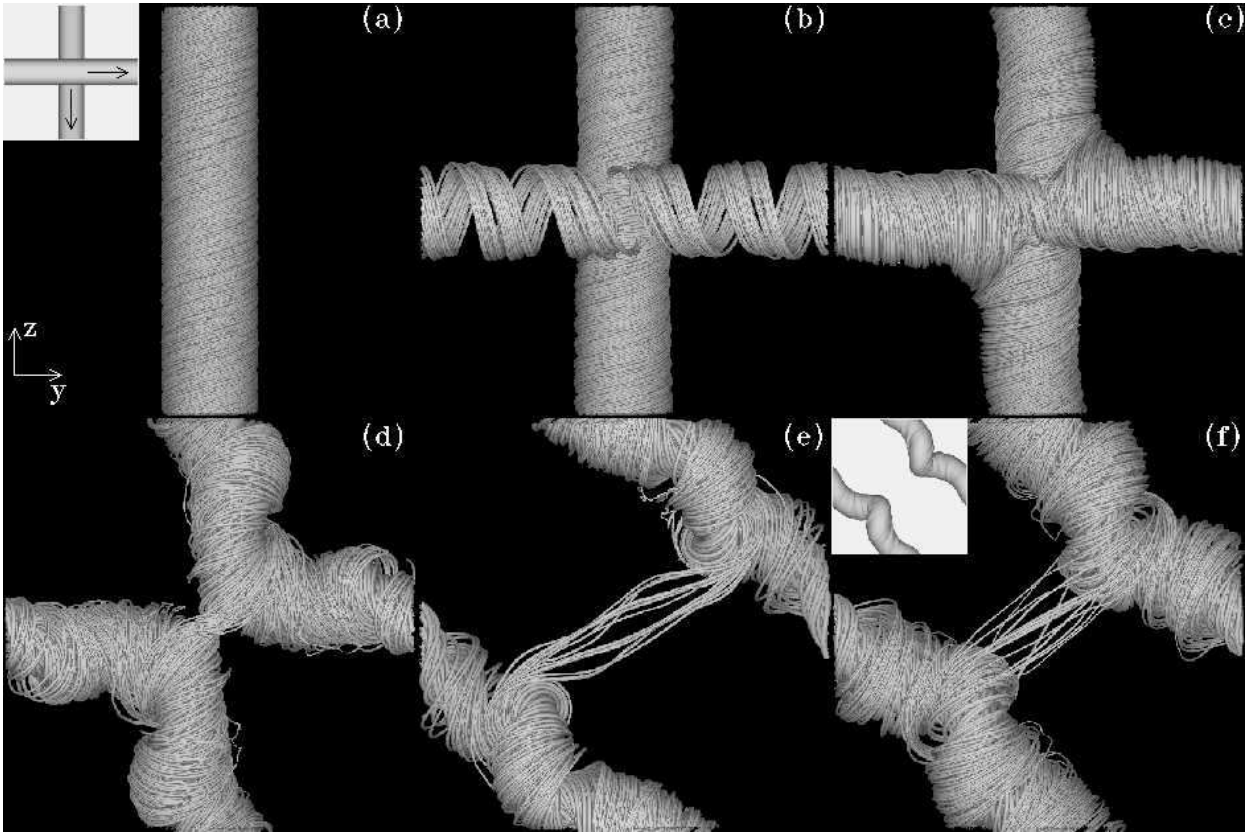}
\caption{Fieldlines from the RR6 simulations at ${\mathcal T}=5$
and for the nonuniform Lundquist number with a maximum
of $288,000$. At this level of twist the tubes slingshot.
The panels shown here are at times $tv_A/R=[0,14,27,45,85,194]$.
\label{fig:rr6q5}}
\end{figure}

\begin{figure}[ht]
\epsscale{1.}
\plotone{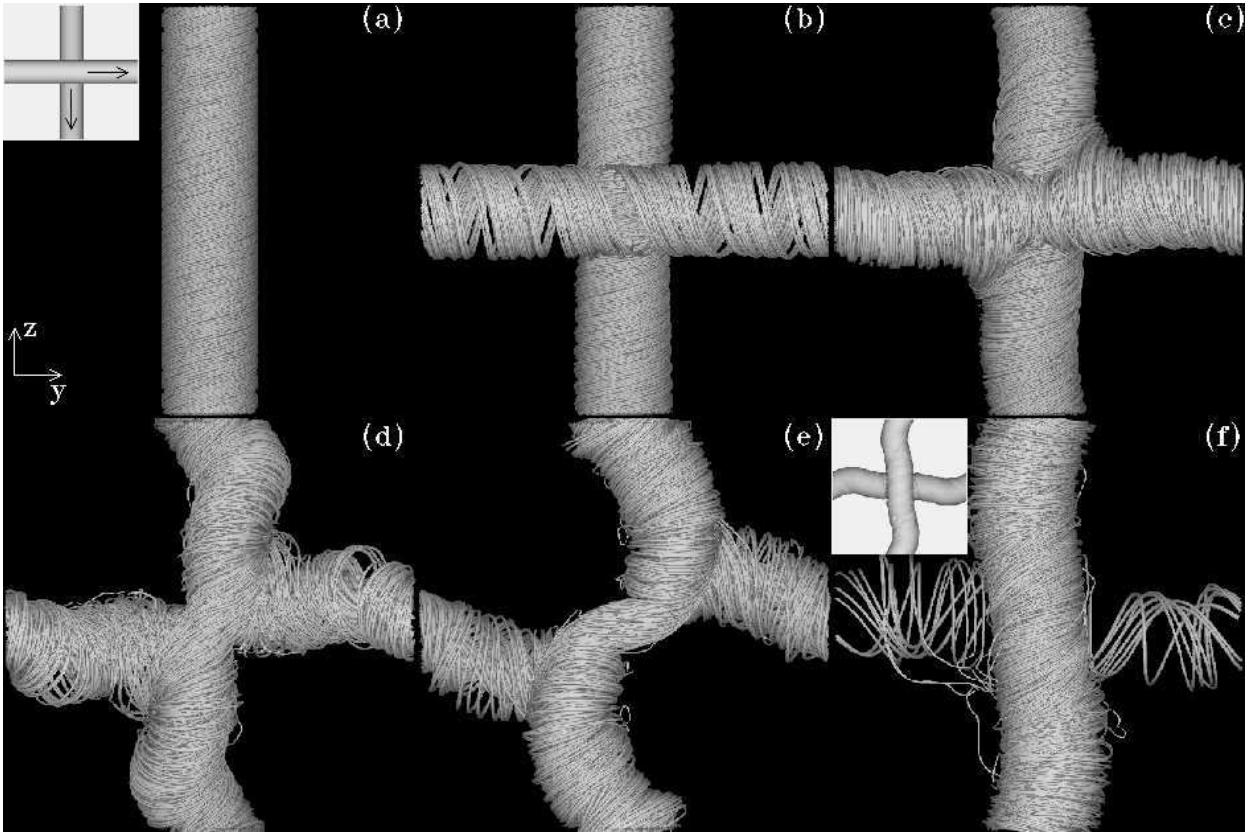}
\caption{Fieldlines from simulation RR6 at ${\mathcal T}=5.5$
and for a nonuniform Lunquist number with a maximum of $288,000$.
At this level of twist the tubes tunnel.
The panels shown here are at times $tv_A/R=[0,14,28,46,79,159]$.
\label{fig:rr6q5.5}}
\end{figure}

\begin{figure}[ht]
\epsscale{1.}
\plotone{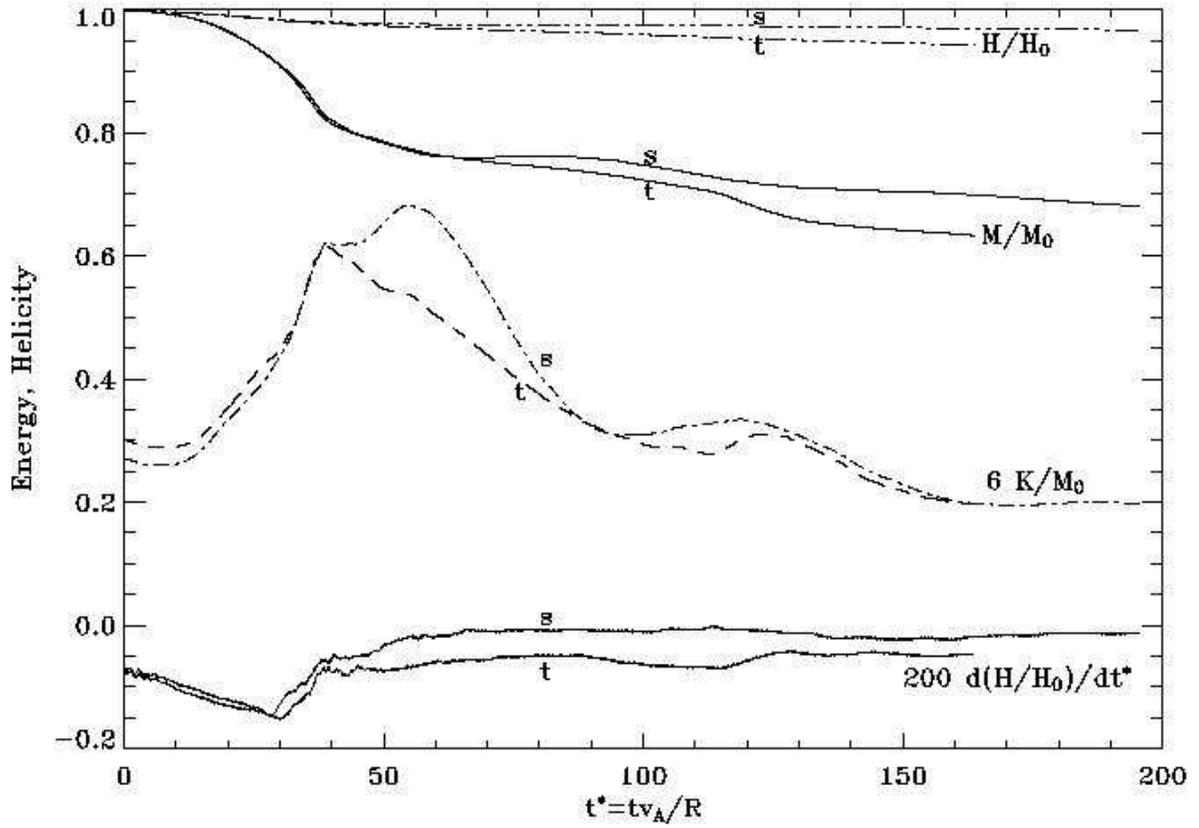}
\caption{Time evolution of helicity, magnetic energy, kinetic energy,
and time derivative of helicity for the RR6 ${\mathcal T}=5.5$ tunnel 
simulation (t) and for the RR6 ${\mathcal T}=5$ slingshot simulation
(s).
\label{fig:rr6energy2}}
\end{figure}

In addition to measuring the twist at which the RR6 interaction
transitions from tunnel to slingshot, we can also
measure the energy release for each interaction, and compare
that with the value predicted by the analytical theory in Figure 
\ref{fig:rr6energy}. By studying the time sequence of the fieldline plots
during the $\tw=5.5$ tunnel reconnection,
we estimate that the slingshot and subsequent rebound reaches
its end state at $tv_A/R=72$, at which point $25\%$ of the magnetic 
energy has been lost.  This can be compared with a prediction by 
the analytical theory of a loss of $12\%$ due to a slingshot. Similarly, at 
$tv_A/R=151$ the tunnel appears to be complete, at which point a total
of $36\%$ of the magnetic energy has been lost to reconnection
and diffusion, compared to a predicted loss of $22\%$ due to
a tunnel reconnection. These values, and the equivalent
values for the other simulations we performed at this
high Lundquist number, are plotted in Figure \ref{fig:rr6energy}
as asterisks for slingshots and as plus signs for the tunnel. 
Note there are both a tunnel and slingshot energy value for the 
$\tw=5.5$ simulation, as we saw both interactions in that case.
As magnetic energy is being released via diffusion
during the entire simulation, as well as via the reconnection,
we expect that more magnetic 
energy will be lost than is predicted by the analytical theory for
reconnection alone. One can see from Figure \ref{fig:rr6energy} that this
is true. The actual energy release for the slingshot
is very close to the predicted value for $\tw=2$ but increases
gradually until about twice the predicted amount is released at
$\tw=5$. This makes sense, as the slingshot
occurs more quickly at low twist, where there is no hint of
a tunnel, than at high twist, where there is a battle between
the slingshot and tunnel which slows the interaction down.
So at high twist, there is more time for the field to lose
energy to diffusion before the interaction is complete.

\begin{figure}[ht]
\epsscale{1.}
\plotone{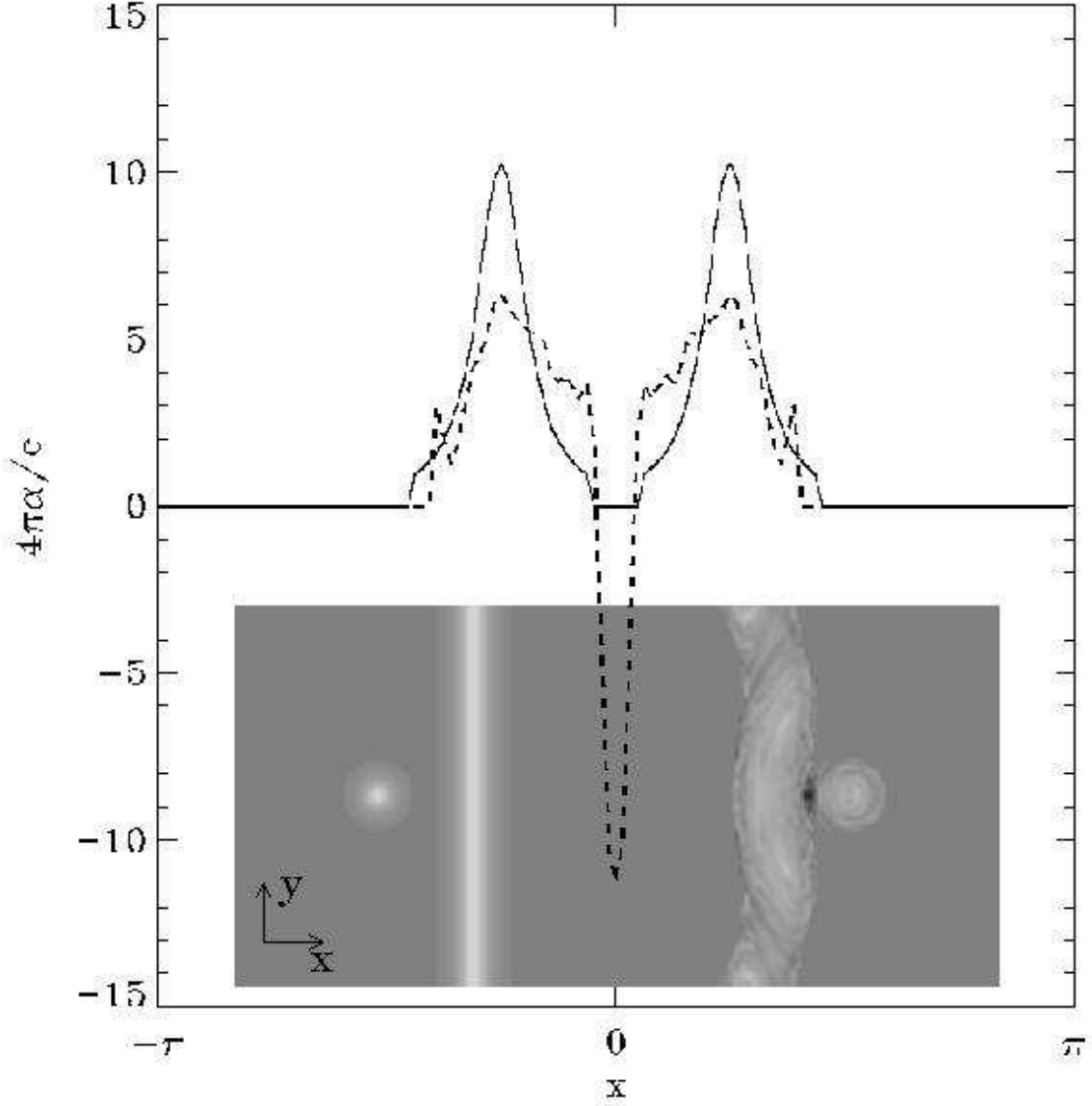}
\caption{Plot of $\alpha$ for the RR6 tunnel simulation of Figure 
\ref{fig:rr6q5.5}. The solid line shows the initial state ($tv_A/R=0$) 
at $y=z=0$, while the dashed line shows the final, tunneled state 
(at $tv_A/R=159$).  The inset greyscale plots show $\alpha$
at the same times in the $z=0$ plane, where the left plot is the
initial state and the right plot is the final state. Here 
black is $4\pi\alpha/c=-15$
and white is $4\pi\alpha/c=15$. This shows that the two
tubes evolve towards constant $\alpha$ fields during their reconnection.
\label{fig:alpha-rr6}}
\end{figure}

A second possible reason for the extra energy release
in the simulations relative to what is predicted by the
theory is that the flux tubes may not evolve homologously
to a uniform twist configuration after reconnection,
but may instead evolve to a lower energy state. \citet{Taylor1986}
showed that the lowest energy state a twisted field can
evolve to while conserving total helicity is a constant $\alpha$
force free state, where
\be
 \alpha\equiv \frac{{\bf J}\cdot{\bf B} }{|{\bf B}|^2}.
\ee
The reconnection which occurs here gives the fields the
opportunity to evolve towards this state.  Figure \ref{fig:alpha-rr6}
shows $\alpha$ for the initial and final states of this simulation,
measured for $|{\bf B}|^2 > |{\bf B}|^2_{max}/10$. 
The line plot shows $\alpha$ along the ${\bf{\hat x}}$ 
axis at $y=z=0$ through the two flux tubes at the start (solid line)
and end (dashed line) of the simulation while the greyscale
plot shows $\alpha$ at $z=0$, with white representing $\alpha=15$
and black representing $\alpha=-15$. The initial, constant
twist state has $4\pi\alpha/c=2\tw/(1+\tw^2r^2)$, and is clearly
not uniform, but the final state is rather more uniform.
A more strict measure of this trend is provided by the standard deviation,
$\sigma$ \citep[see][]{Press1986}, of $\alpha$ over the volume of the 
simulation where $|{\bf B}|^2 > |{\bf B}|^2_{max}/10$. This decreases
from $2.0$ initially to $1.7$ at the end of the simulation. 
The tubes have therefore evolved somewhat closer to the Taylor state,
which may allow them to achieve a lower energy than
the constant twist state we assume in equation (\ref{eq:tunnel}).

\section{ RR5 Reconnection}

We will now study the effect that the flux tube collision
angle has on the tunnel and slingshot reconnections.
The analytical theory shows that the tunnel should occur for any 
pair of positively twisted tubes crossing at an angle 
$\pi < \theta < 2\pi$. At collision angles of $N\pi/4$ this should 
include $N=5$, $6$, and $7$. The RR5 and RR7 interactions were studied
by \citet{Linton2001} at a uniform $S_{\eta}=2880$, but
a definitive tunnel was not seen in either case. The
RR5 interaction bounced with little reconnection,
while the RR7 interaction tunneled partway and then stopped, 
leaving the two tubes entangled \citep[see Fig. 15 of][]{Linton2001}.
As we can now perform these simulations at a higher Lundquist
number, using the localized resistivity to keep the
codes stable, we have revisited these simulations to
see if the tunnel occurs when diffusive loss of twist plays
less of a role.

For a simulation at a background Lundquist number of $288,000$ and 
a collision region Lundquist number of $2880$, we found
that the RR5 tubes simply bounced at high twist, as they did 
in our original simulations for uniform a Lundquist number of 
$2880$.  However, surprisingly, at a lower Lundquist number
of $57,600$ globally and $576$ locally, we found that
the RR5 flux tubes do tunnel. In this case, it appears that a 
$lower$ $L_{\eta}$ in the reconnection region is 
important, at least for a collision speed of 
$.025 v_A$.  Apparently at too high a Lundquist 
number, the tubes bounce before reconnection can take hold.  
This suggests that the tubes might reconnect at higher Lundquist 
numbers if the velocity of collision were lower.  

\begin{figure}[ht]
\epsscale{1.}
\plotone{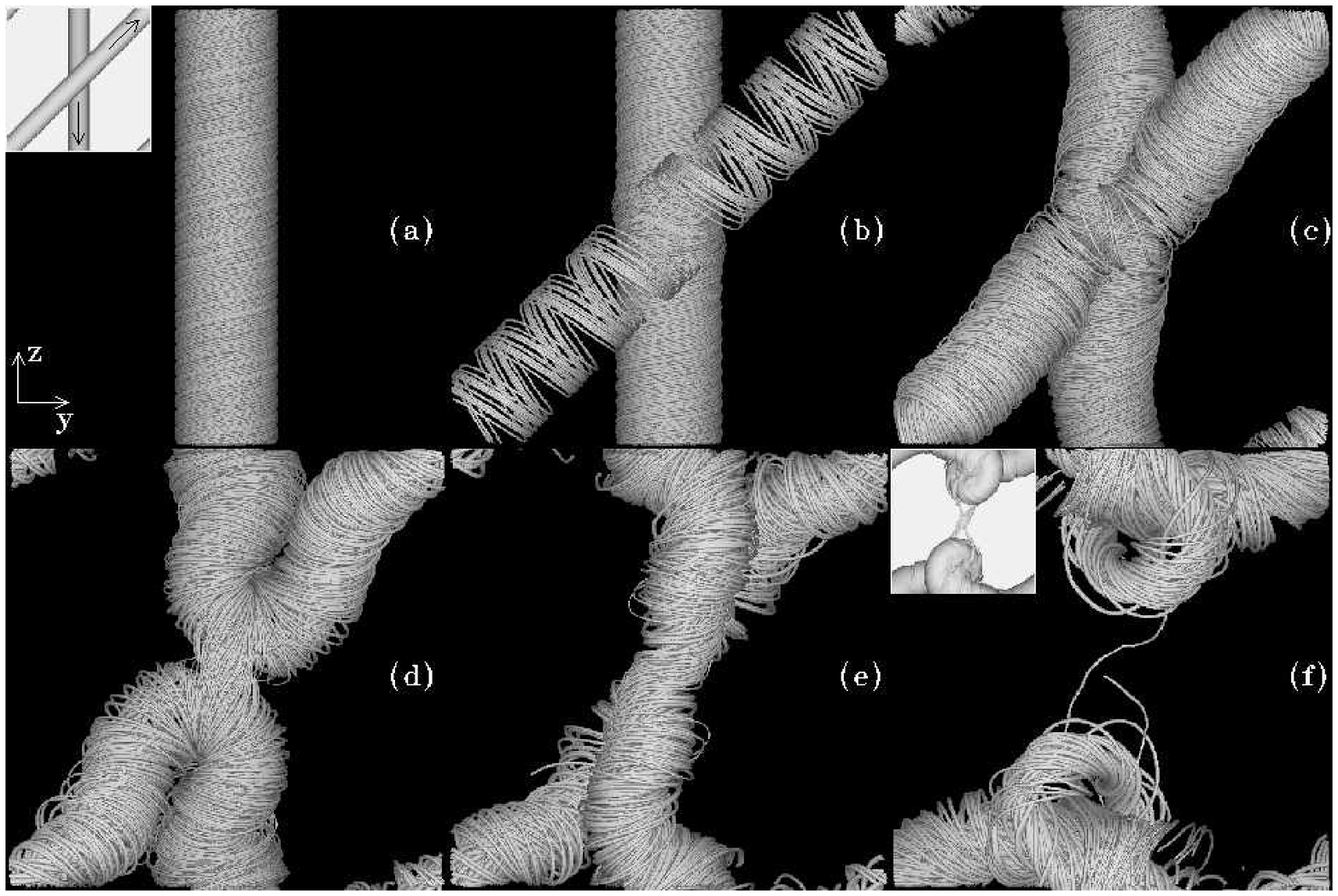}
\caption{Fieldlines from simulation RR5 at ${\mathcal T}=7$
and for a nonuniform Lunquist number with a maximum of $57,600$.
At this level of twist the tubes slingshot.
The panels shown here are at times $tv_A/R=[0,7,32,81,125,248]$.
\label{fig:rr5q7}}
\end{figure}

For these $L_{\eta}=57,600$ simulations, we found a transition from 
tunnel to slingshot at a twist of $\tw=7.5$ to $7$. The $\tw=7$ slingshot
simulation is shown in Figure \ref{fig:rr5q7}. Figure \ref{fig:rr5q7}(a) shows the 
fieldlines of the vertical tube and, in the inset, isosurfaces of both tubes
with arrows indicating the directions of the axial fields.
 The field reconnects from Figures \ref{fig:rr5q7}(b) through 
\ref{fig:rr5q7}(d) to form two $U$ shaped flux tubes which are ready to slingshot 
away from each other to the top and bottom of the simulation. But, as there is
almost enough twist here for the tubes to tunnel, a number of fieldlines
tunnel before the tubes slingshot. These fieldlines briefly dominate the 
interaction so that the field in Figure \ref{fig:rr5q7}(e) looks like it has
almost tunneled. However, this attempt does not succeed, the tunneled fieldlines 
are pulled back, and the tubes slingshot into highly kinked flux tubes by 
Figure \ref{fig:rr5q7}(f). This final slingshot state contrasts with the 
$\tw = 1$ RR6 slingshot of Figure \ref{fig:rr6q1}, where the tubes 
do not kink at all. The difference is that the tubes here are
highly twisted.  While the initial state is only slightly 
kink unstable, the slingshot shortens the tubes significantly,
concentrating their twist, and making them much more kink unstable. 
When the tubes kink, this stretches them to the optimal length
at which the sum of axial and twist field energy is at
a minimum \citep[see][]{Linton1999}.  

\begin{figure}[ht]
\epsscale{1.}
\plotone{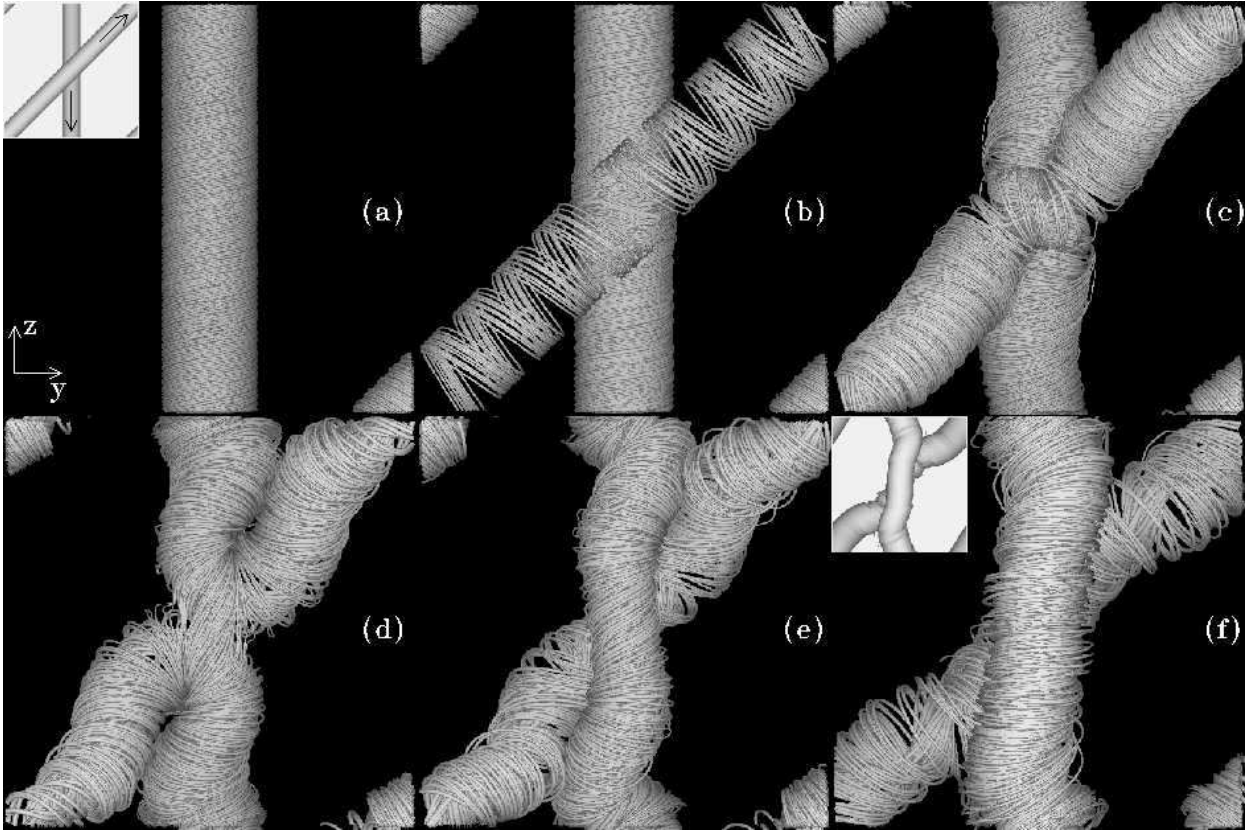}
\caption{Fieldlines from simulation RR5 at ${\mathcal T}=7.5$
and for a nonuniform Lunquist number with a maximum of $57,600$.
At this level of twist the tubes tunnel.
The panels shown here are at times $tv_A/R=[0,7,32,79,106,261]$.
\label{fig:rr5q7.5}}
\end{figure}

The $\tw=7.5$ tunnel simulation is shown for comparison
in Figure \ref{fig:rr5q7.5}. As for the RR6 simulation, the early
part of this reconnection, in Figures \ref{fig:rr5q7.5}(b) through 
\ref{fig:rr5q7.5}(d) looks very much like the corresponding slingshot 
reconnection at lower twist in Figures \ref{fig:rr5q7}(b) through 
\ref{fig:rr5q7}(d). But again, there are more tunneled fieldlines by Figure 
\ref{fig:rr5q7.5}(e) than there are for the slingshot
in Figure \ref{fig:rr5q7}(e). In this case, these fieldines are strong 
enough to dominate the interaction, and so by Figure \ref{fig:rr5q7.5}(f) 
the flux tubes have largely tunneled. This 
is not as clean a tunnel as for the RR6 interaction, as is shown
by the fact that a number of the diagonal fieldlines
have not become disconnected from the vertical flux tube
and disappeared, but still a large part of the flux has 
tunneled. Comparing the isosurfaces inset into Figure \ref{fig:rr5q7.5}(a) and 
\ref{fig:rr5q7.5}(f) gives further evidence of tunneling by showing that the 
tubes have switched positions. The isosurfaces also show that there is significantly more diffusion
in this simulation than in the RR6 simulation at $5$ times the Lundquist
number. The isosurfaces of the RR6 tunneled flux tubes in Figure
\ref{fig:rr6q5.5}(f) are not discernibly thicker than they were initially
in \ref{fig:rr6q5.5}(f). But the isosurfaces of the RR5 tunneled flux
tubes in Figure \ref{fig:rr5q7.5}(f) are $30\%$ to $50\%$ thicker
than they were initially, in Figure \ref{fig:rr5q7.5}(f), indicating
that the tubes have spread out radially due to diffusion.

\begin{figure}[ht]
\epsscale{1.}
\plotone{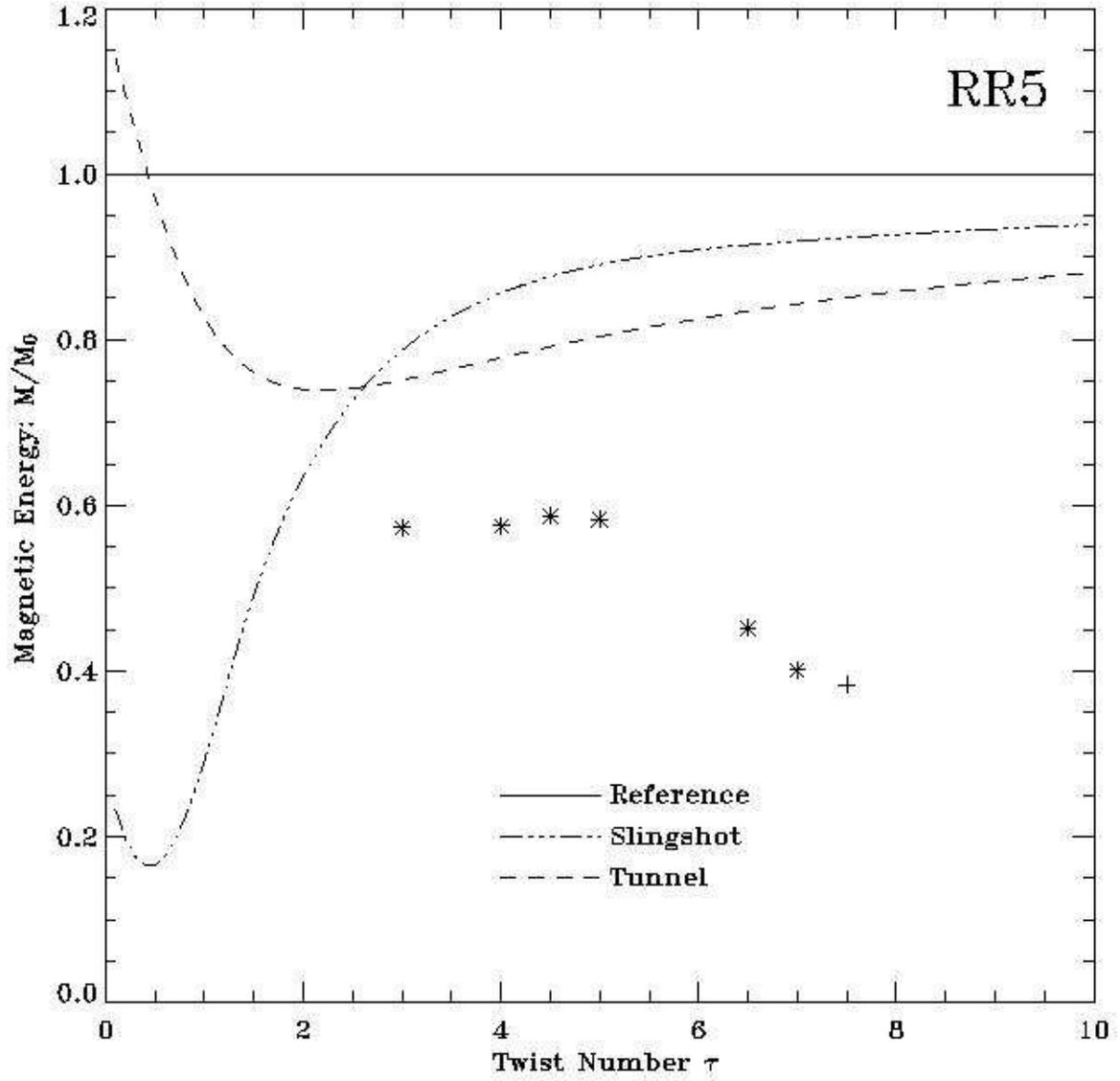}
\caption{
Theoretical reconnection energy, as in Figure \ref{fig:rr6energy}, for RR5 
simulations.  The asterisks denote simulation values for slingshot interactions,
while the plus sign denotes the simulation value for the tunnel 
interaction of Figure \ref{fig:rr5q7.5}.
\label{fig:rr5energy}}
\end{figure}

The energy prediction for RR5 reconnections as a
function of twist is shown in Figure \ref{fig:rr5energy}. This
looks quite similar to that for RR6 interactions,
and level of twist below which the slingshot is predicted to lose more 
energy than the tunnel is about the same, at $\tw=2.62$. 
Thus, while the tunnel does occur at $\theta=5\pi/4$ as predicted, 
the transition from tunnel to slingshot, at $\tw=7.5$ to $7$, occurs
at a twist almost three times larger than predicted by energy 
release alone, at least for this level of resistivity.
This large discrepancy could be partly due to the
lower value of Lundquist number used here: a third
of the helicity is lost to diffusion in the $\tw=7.5$ RR5
simulation, and so we expect an equivalent percentage
of the twist was also lost to diffusion. But the
second, and likely more important effect, is that
it is very difficult for the slingshot flux tubes to
reconnect again once they have lost contact. As
can be seen from Figure \ref{fig:rr5q7}(f), the slingshot
tubes end up very far from each other.  While their 
optimal length is fairly long due to their high twist, they do
not settle into $U$ shaped loops of this length, which might keep
them in contact, but rather settle into kinked loops
of this length. Thus, as in the RR6 simulations,
the tunnel may be a lower energy state at twists
well below $\tw=7$, but it is simply inaccessible
to the tubes, and so does not occur.
For comparison, the estimated energy release of
the various slingshot and tunnel interactions we
simulated are plotted on Figure \ref{fig:rr5energy}. These measured values of
energy release do not agree as well with the predicted values
as they do in the RR6 reconnection, but given the relatively high level 
of resistivity we had to use for these simulations, and the long 
time it took for the interaction to finish, this is only to be expected. 
There is also evidence that these tubes evolve significantly towards
a lower energy constant $\alpha$ Taylor state: the 
initial standard deviation of $\alpha$ is $2.6$ relative to
a standard deviation of $1.5$ after the tunnel.
    
\section{RR7 Reconnection}

\begin{figure}[ht]
\epsscale{1.}
\plotone{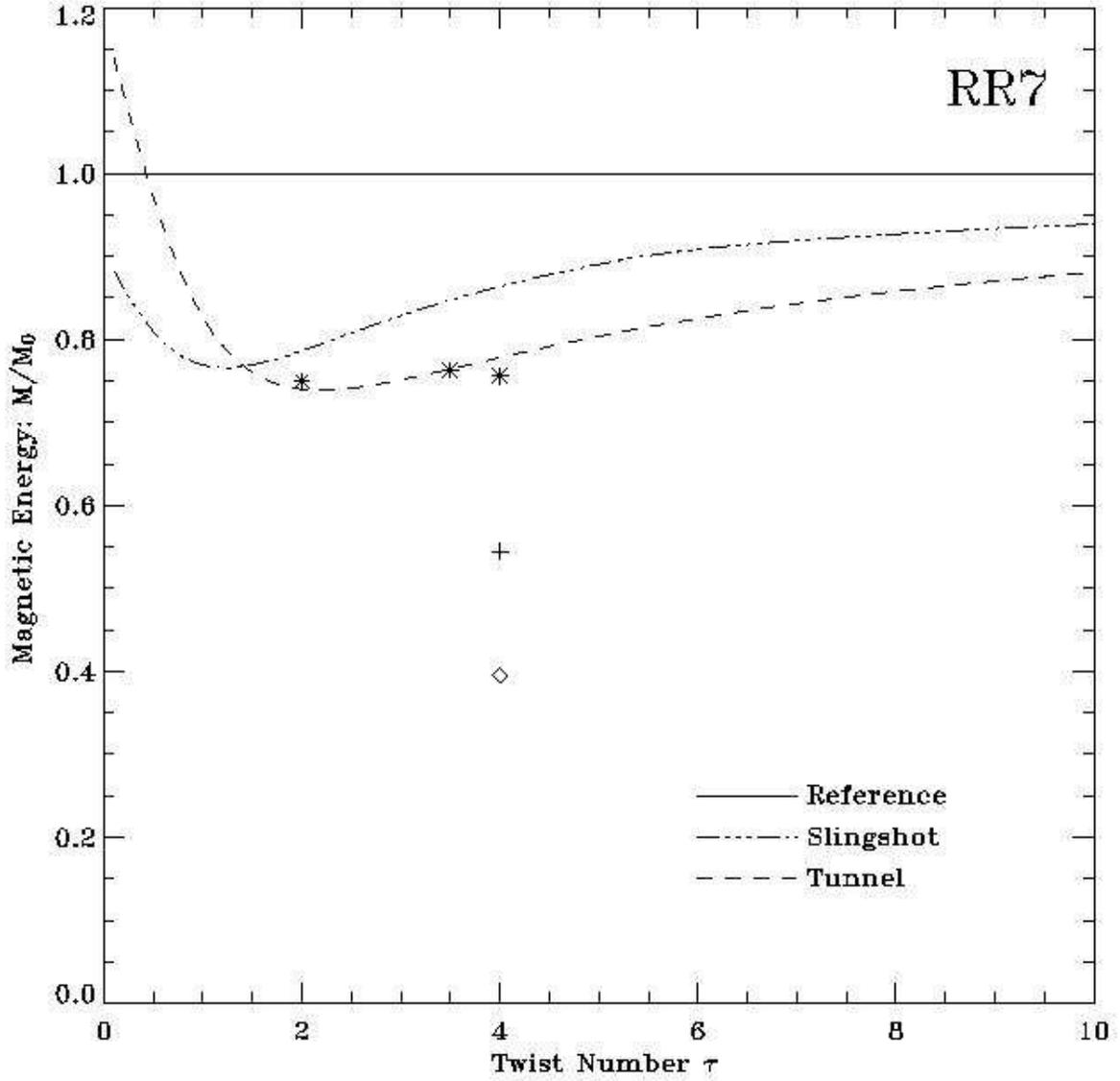}
\caption{
Theoretical reconnection energy, as in Figure \ref{fig:rr6energy}, for RR7 simulations.
The asterisks denote simulation values for slingshot interactions,
while the plus sign denotes the simulation value for the tunnel 
interaction of Figure \ref{fig:rr7q4}(e), and the diamond denotes the
simulation value for the merge interaction of Figure \ref{fig:rr7q4}(f).
\label{fig:rr7energy}}
\end{figure}

\begin{figure}[ht]
\epsscale{1.}
\plotone{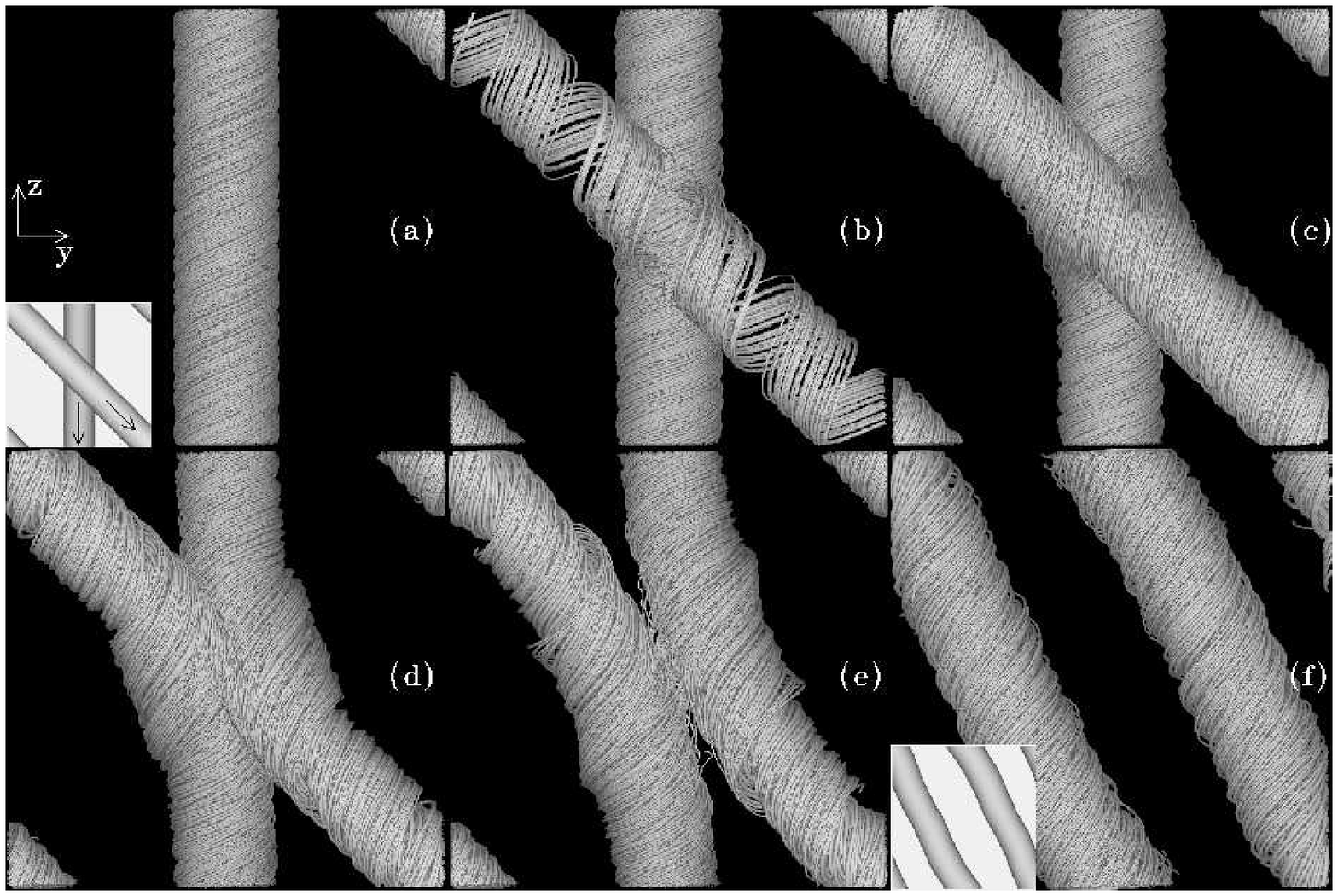}
\caption{Fieldlines from simulation RR7 at ${\mathcal T}=3.5$
and for a nonuniform Lunquist number with a maximum of $57,600$.
At this level of twist the tubes slingshot.
The panels shown here are at times $tv_A/R=[0,7,13,19,25,190]$.
\label{fig:rr7q3.5}}
\end{figure}

For our final set of simulations, we investigated
the reconnection of RR7 flux tubes. Here, we found that the code 
was more unstable than for the RR6 collisions, so we had 
to decrease the nonuniform Lundquist number by a factor of 
$5$ to $57,600$ globally, and $576$ locally, as for the
RR5 simulations.  At this level of resistivity we 
found that there is indeed a clear tunnel interaction, and
that the transition from tunnel to slingshot occurs
at a twist of $\tw=4$ to $3.5$, respectively.
The energy prediction for this interaction, shown in
Figure \ref{fig:rr7energy}, indicates that the slingshot becomes
preferable below twists of about $1.41$, so again the
actual level at which the transition occurs is a factor
of two to three higher than the energy analysis predicts.
Figure \ref{fig:rr7q3.5} shows the $\tw=3.5$ slingshot interaction.
This interaction is quite straightforward, but also
quite interesting in that the angle, $7\pi/4=-\pi/4$, between the
axial field of the flux tubes is so small. Even
at this small angle, the flux tubes reconnect completely
because of the coherence that the twist induces.  The flux tubes 
reconnect from Figures \ref{fig:rr7q3.5}(b) through \ref{fig:rr7q3.5}(e)
to form a simple pair of cylindrical, twisted flux
tubes by the end of the simulation in Figure \ref{fig:rr7q3.5}(f).

\begin{figure}[ht]
\epsscale{1.}
\plotone{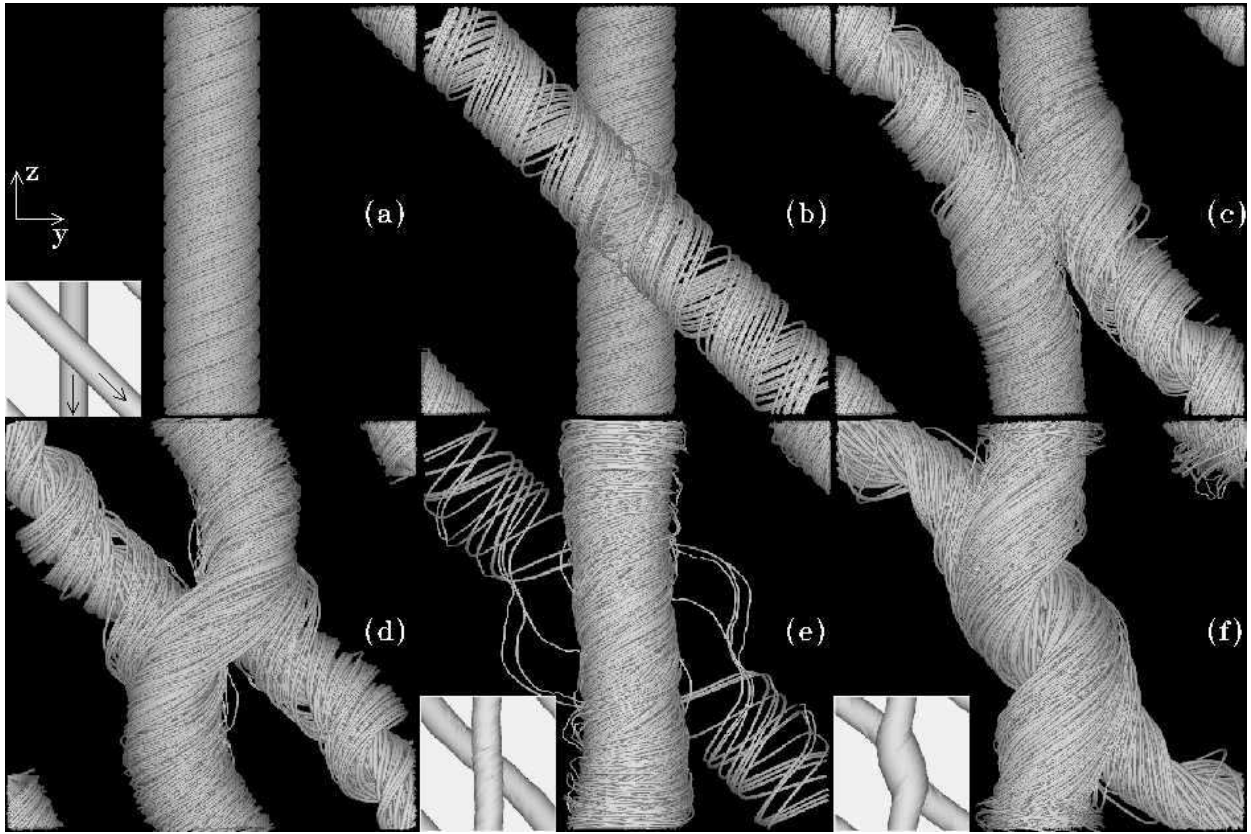}
\caption{Fieldlines from simulation RR7 at ${\mathcal T}=4$
and for a nonuniform Lunquist numberwith a maximum of $57,600$.
This simulation shows three separate reconnection interactions:
the tubes first slingshot by panel (d), then tunnel by panel (e), 
and finally merge together by panel (f).
The panels shown here are at times $tv_A/R=[0,7,25,40,78,202]$.
\label{fig:rr7q4}}
\end{figure}

In contrast, the tunnel reconnection at $\tw=4$, shown in Figure 
\ref{fig:rr7q4} is quite an involved interaction. The tubes start off
by reconnecting to slingshot in Figures \ref{fig:rr7q4}(b) and 
\ref{fig:rr7q4}(c). But they do not completely separate: some of the flux
has already tunneled by the time the tubes try to slingshot away from 
each other, and this keeps them together. The tubes then reconnect a 
second time in Figure \ref{fig:rr7q4}(d) to tunnel by Figure 
\ref{fig:rr7q4}(e). Only a few diagonal fieldlines are still 
connected to the vertical tube at this time, and so these fieldlines 
are mostly invisible. The isosurface inset in Figure \ref{fig:rr7q4}(e) 
presents further evidence of the tunnel when compared with the initial 
isosurface in Figure \ref{fig:rr7q4}(a). But the reconnection does 
not stop here. By tunnelling, the tubes move from an RR7 configuration
to an RR1 configuration. This can be seen by rotating the simulation 
about the ${\bf \hat z}$ axis so that the vertical flux tube is 
behind the horizontal flux tube, as per our convention: in this 
orientation, the tubes cross each other at $\theta=\pi/4$. But this 
is the configuration for which \citet{Linton2001} found that the azimuthal 
component of the field reconnects to merge the two tubes into
a single tube \citep[see Fig. 8 of][]{Linton2001}.  Exactly the same 
thing happens here: the newly tunneled RR1 flux tubes merge together 
to form a single flux tube at the center of the simulation, while 
the boundary conditions keep the footpoints well separated.
This RR7 simulation therefore exhibits three full flux tube reconnections: 
a slingshot, followed by a tunnel,
and then by a merge.  It is now apparent that we saw approximately the
same thing in our RR7, $S_{\eta}=2880$ simulations in \citet{Linton2001}:
due to the faster rate of overall diffusion there, the tubes
started to merge before completing the tunnel, and so the
simulation looked like a failed tunnel interaction.   

Magnetic energy is released by reconnection at each of these three
stages. The estimated measurements of this energy release
are shown on Figure \ref{fig:rr7energy}, 
where the asterisks and the plus sign are the slingshot and tunnel energies, 
respectively, and the diamond is the energy of the merged end state.  
The results for the slingshot at $\tw=3.5$ and a second slingshot at 
$\tw=2$ are also shown.  These energy results are reasonably close to 
the predicted values for the slingshot, probably because the slingshot 
occurs very quickly, and so there is little time for diffusive losses.
Interestingly, in this case the standard deviation of $\alpha$ is
$1.6$ after tunneling versus $1.5$ initially, so the tubes do not appear
to have evolved to a constant $\alpha$ state. Once the
tubes merge, however, the standard deviaton of $\alpha$ has dropped to $1.1$.  
These three tunnel results in combination
provide a convincing argument for revisiting the energy
calculation of \citet{Linton2002} to calculate the final
energy state for flux tubes which evolve to constant $\alpha$
force free states.

\section{Conclusion}

We have studied tunneling reconnection for a variety of 
flux tube collision angles, flux tube twists, and resistive Lundquist 
numbers.  We have compared the results with predictions made
by the analytical theory of \citet{Linton2002}.  The prediction 
that positive twist flux tubes crossing at angles in the range 
$\pi < \theta < 2\pi$ will tunnel for high twist was proved true by 
our simulations: we found tunneling interactions at collision angles of 
$5\pi/4$, $6\pi/4$, and $7\pi/4$. We also found that the tunnel 
transitions to slingshot as twist decreases, as predicted by the 
theory, but that the level of twist at which the transition occurs is 
two to three times higher than predicted.  We hypothesize that this 
is due to the dynamics of the reconnection process, which were not
included in the theory. When flux tubes reconnect the first time to 
slingshot, they spring away from each other and can come to a new
equilibrium wherein they are not in contact with each other. If this 
occurs, the tubes cannot reconnect again to tunnel, even if the 
tunnel would allow them to reach a lower energy state than that of 
the slingshot state.  The magnetic energy released by these interactions 
is reasonably well predicted by the theory, if allowances for diffusive
losses are made. For simulations at very high Lundquist numbers, or 
simulations which progressed quickly and therefore allowed little time 
for energy to diffuse resistively, the predicted reconnection energy 
release is within a factor of two of the energy released in the 
simulation. As resistivity plays a larger role in the dynamics, however, 
more magnetic energy is lost to resistivity, independent of reconnection,
and so there is a larger discrepancy between the simulation and
the theory. In addition to resistive losses, we also find evidence
that the tubes evolve towards constant $\alpha$ equilibria after
reconnection, rather than constant twist as we have assumed:
this would allow for further magnetic energy release. 

For reconnection in the solar corona, where the Lundquist
number is extremely high, diffusive losses should be negligible,
and so the prediction for reconnection energy release should
be in better agreement with the total energy release than
it is in our modest Lundquist number simulations.
The transition from tunnel to slingshot, however, is not expected
to change significantly as the Lundquist number increases,
and so we expect that the actual transition would still be 
$2$ to $3$ times higher than predicted. This means that
only very highly twisted flux tubes are expected to tunnel,
and so this should be relatively rare on the Sun.
It should only occur for the small, highly twisted features
observed within larger scale, less twisted sunspots 
(Canfield, private communication), and for those highly twisted 
convection zone flux tubes which are too twisted to successfully emerge
into the corona, because too much mass is entrained in
their highly twisted loops \citep{Fan2001, Magara2001}.
In addition to these solar applications, this study is
important as a general study of the role that 
helicity conservation plays in reconnection. As we have shown here, 
it can have profound effects, and therefore needs to be taken
into account when modeling three-dimensional reconnection.
In addition, as the slingshot reconnection occurs down
to zero twist, the theory of \citet{Linton2002} as
it relates to slingshot energy release could prove 
useful for predicting energy release in solar interactions.
To explore the true usefulness of these predictions, we
need to modify the theory, first to study constant $\alpha$ final
equilibria, and second to encompass arched magnetic flux
tubes so that the simulations can better match solar conditions. We also need to 
increase the resolution of our simulations so that we can
increase the Lundquist number without inducing numerical 
instability. We will then be able to test whether 
the agreement between the predicted energy release and the 
simulated release continues to improve as the diffusive losses 
decrease.

\acknowledgements

We wish to thank Gunnar Hornig for his derivation of the helicity
calculation method we used for these simulations.  This work was 
supported by NASA and ONR grants.

\eject

%\bibliography{/home/linton/md/mark}

\begin{thebibliography}{}

\bibitem[Antiochos et~al., 1999]{Antiochos1999}
Antiochos, S.~K., DeVore, C.~R., \& Klimchuk, J.~A. 1999,
\newblock ApJ, 510, 485

\bibitem[Berger, 1987]{Berger1987}
Berger, M. 1987,
\newblock JGR, 102, 2637

\bibitem[Berger \& Field, 1984]{BergerF1984}
Berger, M. \& Field, G. 1984,
\newblock JFM, 147, 133

\bibitem[Dahlburg et~al., 1997]{DahlburgAN1997}
Dahlburg, R.~B., Antiochos, S.~K., \& Norton, D. 1997,
\newblock Phys. Rev. E, 56, 2094

\bibitem[Dahlburg \& Norton, 1995]{Dahlburg1995}
Dahlburg, R.~B. \& Norton, D. 1995,
\newblock in Small Scale Structures in Three-Dimensional Hydrodynamic and
  Magnetohydrodynamic Turbulence,  ed. M.~Meneguzzi, A.~Pouquet, \& P.~Sulem,
  (Heidelberg: Springer-Verlag),  317

\bibitem[Fan, 2001]{Fan2001}
Fan, Y. 2001,
\newblock ApJ, 554, L111

\bibitem[Finn \& Antonsen, 1985]{FinnA1985}
Finn, J.~M. \& Antonsen, M.~A. 1985,
\newblock Comments Plasma Phys. Controlled Fusion, 9, 111

\bibitem[Gold \& Hoyle, 1960]{Gold1960}
Gold, T. \& Hoyle, F. 1960,
\newblock MNRAS, 120, 89

\bibitem[Gosling, 1975]{Gosling1975}
Gosling, J.~T. 1975,
\newblock Rev. Geophys. Space Phys., 13, 1053

\bibitem[Hornig, 2005]{Hornig2005}
Hornig, G. 2005,
\newblock in preparation

\bibitem[Linton \& Antiochos, 2002]{Linton2002}
Linton, M.~G. \& Antiochos, S.~K. 2002,
\newblock ApJ, 581, 703

\bibitem[Linton et~al., 2001]{Linton2001}
Linton, M.~G., Dahlburg, R.~B., \& Antiochos, S.~K. 2001,
\newblock ApJ, 553, 905

\bibitem[Linton et~al., 1998]{Linton1998}
Linton, M.~G., Dahlburg, R.~B., Longcope, D.~W., \& Fisher, G.~H. 1998,
\newblock ApJ, 507, 404

\bibitem[Linton et~al., 1999]{Linton1999}
Linton, M.~G., Fisher, G.~H., Dahlburg, R.~B., \& Fan, Y. 1999,
\newblock ApJ, 522, 1190

\bibitem[Linton \& Priest, 2003]{Linton2003}
Linton, M.~G. \& Priest, E.~R. 2003,
\newblock ApJ, 595, 1259

\bibitem[Magara \& Longcope, 2001]{Magara2001}
Magara, T. \& Longcope, D.~W. 2001,
\newblock ApJ, 559, L55

\bibitem[Miki\'c \& Linker, 1994]{Mikic1994}
Miki\'c, Z. \& Linker, J.~A. 1994,
\newblock ApJ, 430, 898

\bibitem[Parker, 1972]{Parker1972}
Parker, E.~N. 1972,
\newblock ApJ, 174, 499

\bibitem[Press et~al., 1986]{Press1986}
Press, W.~H., Flannery, B.~P., Teukolsky, S.~A., \& Vetterling, W.~T. 1986,
\newblock Numerical Recipes: The art of scientific computing,
\newblock (Cambridge: Cambridge University Press)

\bibitem[Priest et~al., 2002]{Priest2002}
Priest, E.~R., Heyvaerts, J.~R., \& Title, A.~M. 2002,
\newblock ApJ, 576, 533

\bibitem[Rosner et~al., 1978]{RosnerTV1978}
Rosner, R., Tucker, W.~H., \& Vaiana, G.~S. 1978,
\newblock ApJ, 200, 643

\bibitem[Schindler et~al., 1988]{Schindler1988}
Schindler, K., Hesse, M., \& Birn, J. 1988,
\newblock JGR, 93, 5547

\bibitem[Schrijver, 1998]{Schrijver1998}
Schrijver, C. J. et~al. 1998,
\newblock Nature, 394, 152

\bibitem[Shibata et~al., 1995]{Shibata1995}
Shibata, K., Masuda, S., Shimojo, M., Hara, H., Yokoyama, T., Tsuneta, S.,
  Kosugi, T., \& Ogawara, Y. 1995,
\newblock \apjl, 451, L83

\bibitem[Taylor, 1986]{Taylor1986}
Taylor, J. 1986,
\newblock Rev. Mod. Phys., 58, 741

\bibitem[Wright \& Berger, 1989]{WrightB1989}
Wright, A.~N. \& Berger, M.~A. 1989,
\newblock JGR, 94, 1295

\end{thebibliography}

\eject

\end{document}